\documentstyle[11pt,aaspp4,psfig]{article}

\def\kms{km~s$^{-1}$}
\def\etal{{\it et al.} }
\def\msun{{\rm M}_{\sun}}

\begin{document}
\hskip 3.truein Accepted: Astronomical Journal
\vskip 50pt
\title{Neutral Gas Distributions and Kinematics \\
of Five
Blue Compact Dwarf Galaxies}
\author{Liese van Zee\footnote{Jansky Fellow}}
\affil{National Radio Astronomy Observatory,\footnote{The National Radio 
Astronomy Observatory is a facility of the National Science Foundation,
operated under a cooperative agreement by Associated Universities Inc.}
 PO Box 0, Socorro, NM 87801}
\affil{lvanzee@nrao.edu}
\authoremail{lvanzee@nrao.edu}
\author{Evan D. Skillman}
\affil{Astronomy Department, University of Minnesota, 116 Church St. SE}
\affil{Minneapolis, MN 55455}
\affil{skillman@astro.spa.umn.edu}
\author{John J. Salzer\footnote{NSF Presidential Faculty Fellow}}
\affil{Astronomy Department, Wesleyan University, Middletown, CT 06459--0123}
\affil{slaz@parcha.astro.wesleyan.edu}
\begin{abstract}
We present the results of high spatial resolution HI observations
of five intrinsically compact dwarf galaxies which are currently experiencing
a strong burst of star formation.  The HI maps indicate that
these systems have  a complex and clumpy interstellar medium.  
Unlike typical dwarf irregular galaxies, these Blue Compact Dwarf (BCD) galaxies 
exhibit strong central concentrations
in their neutral gas distributions which may provide a clue to the origin of
their strong star--burst activity.  Furthermore, 
while all of the systems do appear to be rotating, based on observed velocity
gradients, the kinematics are complex.  All systems have non--ordered kinematic
structure at some level; some of the extended gas is not necessarily
kinematically connected to the main system.  

The observed gas distributions and kinematics place constraints on evolutionary
scenarios for BCDs.  Evolutionary links between BCDs, dwarf irregulars, and
dwarf ellipticals have been postulated to explain their high star formation
rates and low luminosity, low metallicity nature.  The BCDs 
appear to have higher central mass concentrations in both gas and stellar content 
than the dwarf irregulars, indicating that evolutionary scenarios
connecting these two classes will require mass redistribution. In addition,
the fact that BCDs are rotationally supported systems indicates that BCDs are
unlikely to evolve into dwarf ellipticals without substantial loss of angular 
momentum.  Thus, while such evolutionary scenarios may still be possible with 
the aid of mergers or tidal interactions, the isolated nature of BCDs suggests
that the majority of BCDs will not fade to become objects similar to the present 
day dwarf ellipticals.  
\end{abstract}


\section{Introduction}
\label{sec:intro}
Dwarf galaxies provide unique probes
of the star--formation process in low metallicity environments. 
Despite a significant amount of observational and theoretical
work on these systems, we still have many unanswered questions 
about star formation in low mass, low metallicity
objects.  For instance, azimuthal averages
of the HI column density indicate that the gas density in dwarf irregulars
is significantly lower (by factors of 2--3) than the Toomre (1964)
instability criterion throughout the star forming disk
 (e.g., Hunter \& Plummer \markcite{HP96}1996;
van Zee \etal \markcite{vHSB97}1997b; Hunter \etal \markcite{HEB98}1998a). 
 In contrast, preliminary work on Blue Compact Dwarf (BCD) galaxies  suggests that they 
exceed the Toomre instability criterion in the central regions, perhaps
facilitating their high star--formation rates   (e.g., Taylor \etal
\markcite{TBPS94}1994).  While the above studies do not
include the molecular gas component, primarily because it is difficult
to detect in low metallicity systems (e.g., Elmegreen \etal
\markcite{EEM80}1980; Israel \etal \markcite{ITB95}1995),
the neutral gas density appears to have a regulatory role on
the star--formation process in dwarf galaxies.
Thus, to further investigate these regulatory processes, 
we have undertaken high spatial resolution observations of a small
sample of BCDs, systems which are currently undergoing 
vigorous star--formation activity.

The class of objects known as BCDs, or
\ion{H}{2} galaxies, form a heterogeneous sample ranging
from true dwarf galaxies of only 10$^{7}~\msun$ to outlying \ion{H}{2}
regions of moderate sized galaxies.  In this paper, we will use the
term BCD to refer to galaxies which meet the following criteria: (1) the
underlying galaxy which is hosting the strong starburst 
has an absolute blue magnitude fainter than --16; (2) the
optical size is at most a few kpc; and (3) a large fraction of the
optical light comes from a star burst. 

The origin and duration of the starburst phase in BCDs is still unknown. 
One possibility is that the starburst is triggered, either by an interaction
(as discussed by Taylor \etal \markcite{TBS93}1993), or through a
stochastic process.  If the starburst phase in these galaxies is episodic,
there should also exist in the local universe faded, or post--burst, BCDs. 
Several popular evolutionary scenarios suggest that the 3 common types
of dwarf galaxies (dwarf elliptical (dE), dwarf irregular (dI), and BCD) 
are simply 3 different evolutionary
stages for low mass objects (e.g., Dekel \& Silk \markcite{DS86}1986;
Silk \etal \markcite{SWS87}1987; Davies \& Phillipps \markcite{DP88}1988).
Studies of the stellar content and distributions in dwarf galaxies have
thus far been inconclusive in regard to the feasibility of such morphological
transformations (e.g., Loose \& Thuan \markcite{LT86}1986; Drinkwater \& Hardy 
\markcite{DH91}1991; James \markcite{J94}1994; Papaderos \etal \markcite{PLFT96}1996; 
Sung \etal \markcite{SHRCK98}1998).  To date, few {\it kinematic} constraints have
been placed on such evolutionary scenarios, in part due to the dearth of
good quality, high spatial and spectral resolution observations of dwarf galaxies.

This paper presents the HI distribution and kinematics of five 
intrinsically compact galaxies in order to address both the
regulatory role of gas density on star formation activity and
the evolution of starbursting dwarf galaxies. The paper is organized as follows.  
The data acquisition and reduction are presented in Section \ref{sec:data}. 
 The HI distributions and kinematics are discussed in Sections \ref{sec:hi} and \ref{sec:kin},
respectively.  The individual galaxies are described in Section \ref{sec:gals}.
A discussion of the neutral gas density distribution and star--formation threshold is 
presented in Section \ref{sec:thresh}.  In Section \ref{sec:dwfs}, we
discuss the implications on the evolutionary history of BCDs, as derived from the
neutral gas distributions and kinematics.
Our conclusions are summarized in Section \ref{sec:conc}.

\section{Data Acquisition and Analysis}
\label{sec:data}
Multiconfiguration HI observations of five BCDs were obtained with
the Very Large Array\footnote{The Very Large Array
is a facility of the National Radio Astronomy Observatory.}. 
 Complementary optical imaging observations were obtained at
several different telescopes.  In this section, the procedures for
observation and data reduction are discussed.

\subsection{Sample Selection}
To date, few high spatial resolution (5\arcsec~beams) HI synthesis observations
of dwarf galaxies have been  published
(II Zw 40, Brinks \& Klein \markcite{BK88}1988; 
Holmberg II, Puche \etal \markcite{PWBR92}1992; II~Zw~33, Walter \etal \markcite{WBDK97}1997;
I~Zw~18, van Zee \etal \markcite{vWHS98}1998).  We elected, therefore,
to restrict the sample of galaxies to systems which had previously
been observed and detected with the VLA in one of its more compact
configurations (C or D).  These constraints led to the selection of
3 galaxies from the Taylor \etal \markcite{TBGS95}(1995)
sample (UM 439, UM 461, and UM 462), II~Zw~40 (Brinks \& Klein \markcite{BK88}1988), 
and UGC 4483 (Lo \etal \markcite{LSY93}1993) for our first attempt at high spatial
resolution observations of BCD--like objects.  All of the galaxies in the current sample
meet the above criteria for BCDs; while UGC 4483 is not usually considered
a BCD, perhaps due to its proximity, it too is vigorously forming stars
at the present epoch.

The physical properties of the selected galaxies are summarized in Table \ref{tab:global}.
The morphological classification and multiplicity for all systems other
than UGC 4483 were taken from Telles \etal \markcite{TMT97}(1997).    The classification
scheme is based on the lower surface brightness isophotes, where Type I indicates
irregularly shaped isophotes and Type II indicates symmetric isophotes.  Of the
galaxies in the current sample, only II Zw 40 is classified as a Type I BCD.
The multiplicity indicates of the number of knots of star formation; all
of the galaxies in this sample have at least 2 prominent HII regions.  Also listed in 
Table \ref{tab:global} are the assumed distance to each system (based on H$_0$ of 75 
km s$^{-1}$ Mpc$^{-1}$, except for UGC 4483 where the distance is assumed to be the 
same as NGC 2366 (Tolstoy \etal \markcite{TSHM95}1995)), their absolute
blue magnitudes (Stavely--Smith \etal \markcite{SDK92}1992; Telles \etal \markcite{TMT97}1997),
 and oxygen abundances (Melnick \etal \markcite{MTM88}1988; Skillman \etal \markcite{STKGT94}1994).
The absolute blue magnitudes have been corrected for Galactic extinction, but not for
internal extinction or for nebular contributions to the broadband luminosity.
 UGC 4483 is the nearest system in the sample, and
also has the lowest luminosity and metallicity.  The other four galaxies are at comparable
distances (10--15 Mpc) and have moderate luminosities (M$_B$'s of --14 to --16) and
metallicities (1/15 to 1/6 of solar).  Also tabulated are the optical diameters at the 
25th mag arcsec$^{-2}$ isophote, in both arcsec and kpc (de Vaucouleurs \etal \markcite{RC3}1991; 
Salzer \etal \markcite{SESS98}1998a);  these systems are intrinsically small,  
with typical optical sizes of 1--2 kpc.

\subsection{HI Imaging Observations}
\label{sec:hidata}
Observations of II Zw 40, UGC 4483, UM 439, and UM 461/2 
were obtained with the VLA in its B, C, and
CS configurations during 1997. The CS configuration is a modified version of the
C configuration where 2 telescopes from the middle of the east/west arms
are re--positioned at inner D configuration stations to provide short spacings. 
A summary of the observing sessions and the total on--source integration times
are listed in Table \ref{tab:vlaobs}.  
During all observing sessions, the correlator
was used in 2AD mode with the right and left circular polarizations
tuned to the same frequency; the total bandwidth was 1.56 MHz.  The on--line
Hanning smoothing option was selected, producing final spectral data
cubes of 127 channels, each 2.6 \kms~wide.  Standard tasks in AIPS
(Napier \etal \markcite{AIPS}1983)
were employed for calibration and preliminary data reduction.  Each set
of observations was calibrated separately, using 3C 48, 
3C 147, and 3C 286 as flux and bandpass calibrators, and nearby continuum
sources as phase calibrators.  The continuum emission was removed 
with the AIPS task UVLIN prior to combination of uv data sets. 
Only II~Zw~40 had substantial continuum emission associated with the system;
we refer the reader to other studies for discussion of the magnitude and extent 
of this continuum source (e.g., Deeg \etal \markcite{DBDKS93}1993; 
Deeg \etal \markcite{DDB97}1997) since the current observations
were not optimized for continuum detection. 

The line data were transformed to the image plane with several weighting
schemes and combinations of configurations.  To check the data quality,
maps were made for each observing session.  The data cubes presented in this
paper were made from the combined data sets of the B, C, and CS configuration observations.
A robust weighting technique was employed by the AIPS task IMAGR to optimize the beam shape and 
noise levels (Briggs \markcite{B95}1995).  The ``robustness parameter'' in IMAGR controls the
weighting of the uv data, permitting a fine--tuning between sensitivity
and resolution in the final data cubes.  As currently implemented, a robustness
of 5 corresponds to natural weighting of the uv data (maximizing sensitivity
at the cost of spatial resolution) while a robustness of --5 corresponds to
uniform weighting (maximal spatial resolution with lower sensitivity).
Additionally, some maps were made with uv tapers to increase their sensitivity
to low column density material.  The relevant IMAGR parameters for a
selected sample of the maps are listed in Table \ref{tab:maps}.  Throughout
this paper we will refer to the lowest resolution, 
tapered data cubes as the ``tapered cubes'', to the robustness of
5 cubes as the ``natural weight cubes'', and to the robustness of
0.5 cubes as the ``intermediate weight cubes.'' 
Selected channels from the HI data cubes are presented
in Figures \ref{fig:chans40}--\ref{fig:chans462}.    All subsequent
analysis was performed within
the GIPSY package (van der Hulst \etal \markcite{GIPSY}1992).

To determine if the total flux density was recovered in the HI synthesis 
observations, total integrated profiles were constructed from the natural
weight data cubes after correcting for the primary beam shape with the
GIPSY task PBCORR.  For all systems, the total flux density recovered
in the VLA maps is in good agreement (within 5\%)
with previous single dish flux measurements (e.g., Stavely--Smith \etal
\markcite{SDK92}1992; van Zee \etal \markcite{vHG95}1995; van Zee \etal 
\markcite{vMHHR97}1997c; Salzer \etal
\markcite{SRWMB98}1998b; J. Salzer \markcite{S98}1998, private communication).
The total HI masses listed in Table \ref{tab:global}
were calculated from the observed VLA flux density and the assumed distance to
the system.   

Moment maps of each data cube were computed in the following manner.
For each galaxy, the natural weight cube was smoothed to a resolution of twice 
the beam; the smoothed cubes were clipped at the 2 sigma level (the noise level was
measured in several line--free regions of the smoothed cube); the resultant
clipped cube was then interactively blanked to remove spurious noise spikes.
Signal was identified based on spatial continuity between channels.   A
conditional transfer was used to blank the corresponding locations in both
the natural and intermediate weight data cubes (corrected for primary
beam attenuation) based on the blanked, smoothed
cube.  Moment maps of the smoothed, natural, and intermediate weight data 
cubes were created with the GIPSY task MOMENTS and are presented in 
Figures \ref{fig:iizw40}--\ref{fig:um462}.

\subsection{Optical Imaging}

Optical broad band and H$\alpha$ images were obtained during several
observing runs.  The observing parameters
are summarized in Table \ref{tab:optobs}. 
 The observation and reduction procedures for the H$\alpha$ image 
of UGC 4483 are described in Skillman \etal \markcite{STKGT94}(1994).
H$\alpha$ and B--band images of UM 439, UM 461, UM 462, and II Zw 40
were obtained with the KPNO No. 1 0.9m telescope and followed similar
reduction procedures as those described in Salzer \etal \markcite{SAMGH91}(1991).
Finally, R--band images of II Zw 40 and UGC 4483 were obtained with the
KPNO 0.9m telescope; the observation and reduction procedures for these images
were identical to those described in van Zee \etal \markcite{vHS97}(1997a).
Plate solutions for the optical images were derived from coordinates
of at least five stars listed in the APM catalog (Maddox \etal \markcite{MSEL90}1990)
and are accurate to 0.5\arcsec.

	H$\alpha$ luminosities were computed from the observed H$\alpha$ flux
and the assumed distance to each system.  The
H$\alpha$ luminosity was then converted to the current star--formation rate (SFR),
assuming a Salpeter IMF
(Hunter \& Gallagher \markcite{HG86}1986):
\begin{equation}
{\rm SFR} = 7.07 \times 10^{-42}~{\rm L(H\alpha)~M_{\odot}~yr^{-1}}.
\end{equation}
The current SFR and the gas depletion time scale (M$_{\rm HI}$/SFR) are tabulated in 
Table \ref{tab:global}.  Note that if the IMF is biased towards high mass stars, the 
calculated gas depletion time scale will be an underestimate.  The derived gas depletion
time scales should be indicative, however, since observations of the stellar populations
in starbursting (e.g., R136, Massey \& Hunter \markcite{MH98}1998) and low metallicity 
environments (e.g., SMC and LMC, Massey \etal \markcite{MLDG95}1995), are consistent 
with a Salpeter IMF. In general, the current SFRs of the BCDs are quite high, with 
typical gas depletion times of approximately 1 Gyr.
  
\section{HI Morphology}
\label{sec:hi}

A significant fraction of the gas associated with dwarf galaxies
frequently extends well beyond the optical system (e.g., 
DDO 154, Carignan \& Beaulieu \markcite{CB89}1989; NGC 4449,
Bajaja \etal \markcite{BHK94}1994; Hunter \etal \markcite{HWvGK98}1998b).
The HI diameters of the BCDs (defined as the maximum diameter measured at
a column density of 10$^{20}$ atoms cm$^{-2}$) were obtained from the 
HI column density maps made from the low resolution tapered data cubes
[Figures \ref{fig:iizw40}(a)--\ref{fig:um462}(a)]. 
The HI--to--optical diameter ratios are listed in Table \ref{tab:global}.  
The HI distributions of all five systems are extended,
with sizes typically 3--4 times that of the optical diameter.
II Zw 40 is the most extreme, with tidal tails extending approximately 8.5\arcmin~in
diameter (see Section \ref{sec:iizw40});  the HI diameter listed in 
Table \ref{tab:global} for this galaxy was obtained by measuring only the 
inner HI distribution, along a position angle of 115\arcdeg.
These results are consistent with other spatially resolved observations of 
BCD--type galaxies, which have also
found large HI--to--optical diameter ratios 
(e.g., Taylor \etal \markcite{TBPS94}1994; van Zee \etal \markcite{vWHS98}1998).  

The wealth of detail visible in the high
spatial resolution maps is striking.  
Figures \ref{fig:iizw40}(b)--\ref{fig:um462}(b)
illustrate the clumpiness of the ISM on 10\arcsec~spatial scales; even
more detail is seen at the highest spatial resolution of these observations
[Figures \ref{fig:iizw40}(c)--\ref{fig:um462}(c)].  For instance, the
HI distribution for UGC 4483 (linear scales of 100 to 300 pc) is 
clearly clumpy on all spatial scales sampled.  

Inspection of the outer HI distributions reveal varying degrees of irregularity
in both the column density distribution and kinematic continuity.  For
example, in UM 439 there appears to be an associated HI cloud to the east 
of the main galaxy.  This cloud is likely to be a real feature since it
was detected in both the C and CS configuration observations,
as well as in previous D configuration observations (Taylor \etal \markcite{TBGS95}1995).
Another example of this phenomenon is the gas to the north of UM 461 which
appears to be kinematically distinct in the channel maps (Figure \ref{fig:chans461})
but is not clearly spatially separated from the main galaxy.  Similar
HI clouds have been detected around other star forming galaxies (e.g., Taylor
\etal \markcite{TBGS95}1995; Wilcots \etal \markcite{WLM96}1996).  However,
not all low mass galaxies have such features  (e.g., van Zee \etal \markcite{vHSB97}1997b).
The source and ultimate fate of the gas in the periphery of these galaxies is
unclear.  Several possible origins are: (1) warping of the outer gas disk; (2) 
ejecta from previous episodes of star formation; (3) material still in the
process of accretion; or (4) tidal debris.  The present observations of the
gas kinematics and distribution cannot constrain the origin of this outlying material.

Azimuthal averages of the HI surface density of the BCDs are shown in
Figure \ref{fig:radprofs}.  In addition, comparison samples of dIs
(van Zee \etal \markcite{vHSB97}1997b) and a few additional BCDs
(Taylor \etal \markcite{TBPS94}1994) have been taken from the literature.
As previously remarked by Taylor \etal (\markcite{TBPS94}1994; \markcite{TBGS95}1995), 
BCDs tend to have centrally peaked surface densities.  In contrast,
dIs tend to have flat HI distributions in the inner regions, more similar to 
the HI distributions of spiral galaxies (e.g., Cayatte \etal \markcite{CKBv94}1994).
Not only are the BCDs centrally concentrated, their central surface densities are
significantly higher than those of the dIs.
Analysis of their underlying stellar light distributions (i.e., with the starburst 
removed) also indicates  that BCD host galaxies  have more
centrally concentrated stellar mass densities than dIs (e.g., Norton \& Salzer 
\markcite{NS98}1998).  Thus, the BCDs appear to have inherently different stellar 
and gas mass distributions than dIs.  This may
be a clue to the origin of strong star--formation activity in BCDs.

\section{ Kinematics}
\label{sec:kin}
In this section we present the results of kinematic studies based on the
HI velocity information.  The observed
velocity gradients are used to estimate dynamical masses in Section \ref{sec:vel}.
In Section \ref{sec:disp}, we demonstrate that these systems are dominated by rotation. 

\subsection{Velocity Gradients}
\label{sec:vel}
The velocity fields (Moment 1) were calculated from the blanked natural
weight data cubes and are presented in Figures \ref{fig:iizw40}(d)--\ref{fig:um462}(d).
The velocity fields are complex.  II~Zw~40  has extensive tidal tails, with
velocity continuity from the main body to the ends of the tails.  In this system,
the southeastern tail has a reversal in the velocity field, suggesting that the most
extended material is starting to fall back onto the main galaxy (cf. Hibbard \& Mihos
\markcite{HM95}1995) (see Section \ref{sec:iizw40} for further discussion of this system).   
While none of the other galaxies have apparent tidal features, other kinematic irregularities 
such as warps (e.g., UGC 4483) and kinematically distinct clumps of HI gas (e.g., UM 439)
are present in all the systems.   Nonetheless, despite their complexity, all  have 
velocity gradients across the galaxy.  Inspection of the channel maps 
(Figures \ref{fig:chans40}--\ref{fig:chans462}) indicates a clear axis of rotation 
for each of the galaxies. 

The complexity present in the velocity fields  inhibits dynamical 
investigations using standard tilted--ring models.  Nonetheless, we have 
obtained rough estimates of the dynamical masses for 3 of the 5 systems 
based on the observed velocity gradients and estimates of the inclination 
angle.  In standard tilted--ring analyses, the center coordinates, 
systemic velocity, position angle, inclination angle, and rotation velocity 
are fit independently for each ring.  In our analysis, we required that all 
parameters, except the rotation velocity, be constant for all rings.
Further, the inclination angles were fixed based on the observed HI isophotal axial
ratios rather than on kinematic constraints.  The inclination estimates assumed
a thin gas disk; if the gas disk is thick, as may be the case for low mass
galaxies (e.g., Puche \etal \markcite{PWBR92}1992), the inclination will be larger
than estimated here. The derived parameters
are tabulated in Table \ref{tab:hiparms}.  Tilted--ring models were not attempted
for II~Zw~40 or UM 461; their tabulated parameters are rough estimates at best.

  Position--velocity diagrams with the derived rotation curves superposed are
shown in Figure \ref{fig:pv}.  The natural weight data cubes were sliced along
the nominal position angles, centered at the kinematic center.  As frequently
found in low mass systems, the position--velocity diagrams are consistent
with solid body rotation 
throughout the galaxy. The outermost points of the rotation curves (R$_{\rm max}$ 
and V$(R_{\rm max})$) are tabulated in Table \ref{tab:hiparms}.
 Dynamical masses, based on the outermost points of
the rotation curves, are also tabulated in Table \ref{tab:hiparms};  
we emphasize that these masses were derived from only rough dynamical estimates.
The ratio of M$_{\rm HI}$/M$_{\rm dyn}$, typically found to be approximately 0.1
for dwarf galaxies (e.g., Skillman \etal \markcite{SBMW87}1987; 
van Zee \etal \markcite{vHSB97}1997b), is approximately 0.3 for
these BCDs (Table \ref{tab:hiparms}).  Whether these high values of
M$_{\rm HI}$/M$_{\rm dyn}$ are significant
is unclear; the kinematic models were relatively unconstrained, so
it is possible that the dynamical mass has  been underestimated
in these calculations.  Interestingly, Salzer \etal \markcite{SRWMB98}(1998b)
also find higher than typical M$_{\rm HI}$/M$_{\rm dyn}$ for BCDs, based on 
single dish HI spectra.  On the other hand, other spatially resolved observations of
BCDS (e.g., Taylor \etal \markcite{TBPS94}1994) have resulted in values of 
M$_{\rm HI}$/M$_{\rm dyn}$ more typical of dwarf irregular galaxies.  
Further high spatial and spectral resolution
observations of BCDs will be needed to determine if a large fraction
of their total mass is usually in the form of neutral hydrogen.

\subsection{Thermal Motions}
\label{sec:disp}

Gaussian fits to spectral profiles from the natural weight data cubes
were used to determine the average velocity dispersion for each galaxy.
The median $\sigma$ is tabulated in Table \ref{tab:hiparms}.
The observed velocity dispersions
are slightly higher than typically found in dwarf galaxies
(e.g.,  Young \& Lo \markcite{YL96}1996; van Zee \etal \markcite{vHSB97}1997b), 
but this may be due to the moderate velocity gradients and compact nature of
these galaxies resulting in an additional broadening component due to rotation
within the beam.   Unfortunately, due to their lower signal--to--noise,
velocity dispersions could not be measured
accurately from the higher spatial resolution maps.
  Nonetheless, the observed velocity dispersions are
significantly less than the rotational component in these galaxies.  Thus,
these systems are {\it rotation dominated}.

\section{Individual Galaxy Characteristics}
\label{sec:gals}
In this section we present qualitative descriptions of
the individual galaxies.  The new observations are also
compared with previous data from the literature.

\subsection{II Zw 40}
\label{sec:iizw40}
II Zw 40 (UGCA 116) is one of the prototypical BCDs (e.g., Sargent \& Searle 
\markcite{SS70}1970; Searle \& Sargent \markcite{SS72}1972).  It is the 
only galaxy in the present sample classified as a Type I BCD (Telles \etal 
\markcite{TMT97}1997), and is also the only galaxy in this sample that 
is clearly the result of a merger.  The possibility that the burst of 
star formation in II~Zw~40 was the result of an interaction was first 
raised by Baldwin \etal \markcite{BST82}(1982), based on the optical 
morphology.  HI imaging observations confirmed the merger scenario for 
this galaxy and led to speculation that tidal interactions might
be necessary to trigger the high star--formation rates in BCDs
(Brinks \& Klein \markcite{BK88}1988).  

The combination of several
pieces of evidence suggests that  II~Zw~40 is likely to be a merger, not 
merely the result of tidal interactions: (1)  the optical 
morphology is consistent with that of two overlapping/merging dwarf galaxies, 
(2) there is no evidence for a second optical galaxy anywhere along the HI
tidal tails, and (3) a strong color gradient indicates that the low
surface brightness optical extensions are from an older stellar
population.  In particular, the LSB extension to the
south of the intense star burst is quite red, indicating that these
stars are from a much older population than the current starburst.  The
colors (corrected for Galactic reddening and nebular emission) range 
from B--H $\sim$ 1.0 at the starburst to B--H $\sim$ 3.1 in
the LSB southern arm (Salzer \etal \markcite{SESS98}1998a).  The
extension to the SE is intermediate
in color, and also exhibits a modest level of H$\alpha$ emission,
indicating that star formation may be occurring there as well.  The morphology of
the system, as well as the extreme color gradients, are consistent
with the picture that II~Zw~40 is an ongoing merger of two gas--rich
dwarf galaxies.

The new high sensitivity 
HI maps presented here indicate that the tidal tails are
even more extensive than illustrated in Brinks \& Klein \markcite{BK88}(1988). 
In many regards, the HI distribution and velocity field
in II Zw 40 is reminiscent of NGC 7252, a member of Arp's
\markcite{A66}(1966) atlas which has been designated as a prototypical
merging pair of galaxies by Toomre \markcite{T77}(1977).  The HI tidal
tails are drawn out from the optical galaxies, and  
a reversal in the velocity gradient is detected
in the southeastern tail.  While the masses of the components
of the NGC 7252 merger are an order of magnitude higher than those
of II Zw 40 (cf.\ Hibbard \etal \markcite{HGvS94}1994), it is interesting to
compare these two systems.  Hibbard \& Mihos \markcite{HM95}(1995) conducted 
N-body simulations of NGC 7252 and determined that the
velocity reversals seen in the tidal tails of that system were
due to material falling back towards the dynamical center.
In both NGC 7252 and II Zw 40, the radial velocities at the ends
of the tidal tails are similar to the radial velocities of
the main bodies.  The simulations of Hibbard \& Mihos \markcite{HM95}(1995) 
give a detailed timetable for the return of tidal material to
the system.  The return is spread out over several Gyr, but
more than half of the gas has returned in the first Gyr.
Without a similar detailed model of II Zw 40, quantitative
estimates of the fallback of material to the system is not
possible (and we would encourage others to carry-out such 
modelling).  With such modelling, it will be possible to 
determine if infall could sustain star formation at an elevated
rate and increase the lifetime of the burst or result in a
second episode of increased star formation.  Since $\sim$40\% of the gas is
in the tidal tails, the return of this gas to the main body
could have a significant effect on the future star formation 
of the system.   A modelling investigation of the
low mass merger  II Zw 40 could provide valuable insight into the future
evolution of this system.

\subsection{UGC 4483}
\label{sec:u4483}
UGC 4483,  normally considered an Im galaxy, was included 
in the BCD sample because of its compact nature and moderate star--formation
rate.  If this system were placed at a distance of 10 Mpc (the typical 
distance of the other galaxies in this sample), it might be classified as a BCD.
As mentioned previously, the BCD class is quite heterogeneous; UGC 4483
would qualify as a weak BCD in that its star forming regions contribute
a large fraction of its total luminosity.  On the other hand, its gas depletion 
time scale is fairly long, almost 10 Gyr, similar to that found for more
quiescent dwarf irregulars (e.g., van Zee \etal \markcite{vHS97}1997a).

In this system, several knots of star formation are superimposed on a low
surface brightness stellar component.  The brightest knot of
active star formation is located to the north of the optical center
and is associated with the peak neutral gas column density.  As previously
noted by Lo \etal \markcite{LSY93}(1993), the HI
distribution extends well beyond the optical system.  

This galaxy is the nearest object in the sample and thus the HI
observations have the highest linear resolution, on the order of
100 pc beam$^{-1}$.  On these small spatial scales, the HI distribution 
is distinctly clumpy and irregular [Figure \ref{fig:u4483}(b)].
Interestingly, despite the high linear resolution of these observations,
there are no obvious HI holes, such as those seen in IC~10 (Shostak \& Skillman 
markcite{SS89}1989), Holmberg II (Puche \etal \markcite{PWBR92}1992), and
VV 794 (Taylor \etal \markcite{TBPS94}1994).  Nonetheless, at this resolution
a strong correlation is seen between column density peaks and 
sites of active star formation.  Furthermore, this system allows
an estimate of the effect of beam smearing on the observed peak 
column densities.  As expected, low resolution observations have 
correspondingly lower peak column densities (Table \ref{tab:colden}).  
For instance, an increase in the spatial resolution by a factor of 5 
corresponds to an increase in the observed column density by a factor of 2.

\subsection{UM 439}
\label{sec:um439}
The most intense region of star formation is located south of the
center of the optical portion of UM 439 (UGC 6578), and the
highest column density of HI gas occurs at this location as well.  There
is weaker, more diffuse, H$\alpha$ emission over most of the central
part of the galaxy, suggesting that there has been vigorous star
formation occurring there in the recent past.  This activity is
apparently dying out.  The location of the strong knot on the outskirts
of the optical galaxy implies an outward propagation of star formation.
Despite the appearance of recent strong activity in the central portion of
the galaxy, there is no evidence for an HI hole in the middle of this 
object as is seen in other dwarf galaxies (Puche \& Westpfahl \markcite{PW94}1994). 
At the distance of UM 439, however, the detection of such holes
would be difficult.  The 
neutral gas is centrally peaked, and extends well beyond the star 
forming regions.  Previous HI mapping of this
galaxy indicated the presence of extended gas to the
east (van Zee \etal \markcite{vHG95}1995; Taylor \etal 
\markcite{TBGS95}1995).  The new observations confirm the
extended structure to the east, but the total mass of this object is
only $\sim$6 $\times$ 10$^{6}~\msun$, approximately 2\% of the total
HI in this system.  Given its low mass and low column density, it
is unlikely to be a companion to UM 439.  The present observations
are not able to determine the origin of this gas cloud, but it would
be of great interest to determine if this extended material has been
significantly enriched by previous star formation episodes.

\subsection{UM 461 and UM 462}
\label{sec:um461/2}
Two bright knots of active star formation dominate UM 461.
The HI distribution extends well beyond the optical system.
Like UM 439, the extended gas on the east side of UM 461 is kinematically
complex.  The extended structure visible to the SE in
the Taylor \etal \markcite{TBGS95}(1995) maps is not recovered here; 
this feature was probably introduced by solar interference in
their D configuration maps (Taylor \etal \markcite{TBGS96}1996).  
The velocity field [Figure \ref{fig:um461}(d)] 
indicates that there is a velocity gradient across this system,
but the field is too complex to permit rotation curve fits. 

UM 462 (UGC 6850; Mk 1307) is dominated by two knots of star formation,
the brighter of which is centrally located in the galaxy.  The
HI distribution extends well beyond the optical system and peaks
near the central star forming regions.  The velocity field is
indicative of solid body rotation.  

In contrast to the image presented in Taylor \etal \markcite{TBGS95}(1995), 
UM 461 and UM 462 do not appear to be tidally interacting (Figure \ref{fig:um461/2}). 
 Assuming that the systems are both at a distance of 13.9 Mpc, the linear separation
between the two systems is approximately 70 kpc (40 times the
average of their optical diameters, and 11 times their mean HI diameters).
Assuming a relative velocity of $\sim$100 \kms, the crossing time
is 7 $\times$ 10$^{8}$ years, significantly longer than the age of the
star bursts (expected to be less than 100 Myr).  Furthermore, both UM 461 and
 UM 462 are classified as Type II BCDs by Telles \etal \markcite{TMT97}(1997), 
indicating that they are symmetric objects in the optical.  
Thus, even if these systems are physically associated, it is unlikely that they
induced the high rates of star formation in each other.  

\section{Gas Density Thresholds}
\label{sec:thresh}

As discussed in Section \ref{sec:hi}, BCDs tend to have 
higher central gas mass concentrations than dIs, suggesting
an intrinsic difference in the HI distribution between
quiescent dIs and BCDs.  Previous studies of the neutral
gas distribution in BCDs and dIs indicate that the 
gas column density can play a critical role in regulating
star--formation activity (e.g., Skillman \etal \markcite{SBMW87}1987;
Taylor \etal \markcite{TBPS94}1994; Hunter \& Plummer \markcite{HP96}1996; 
van Zee \etal \markcite{vHSB97}1997b; Hunter \etal \markcite{HEB98}1998a).
In studies such as these, the gas density should include both the
atomic and molecular gas components.  In practice, however, the
molecular gas distribution is usually unknown due to the difficulty
detecting CO in low mass, low metallicity galaxies (e.g., Elmegreen \etal
\markcite{EEM80}1980).  Furthermore, even when CO is detected, the conversion
factor from CO to H$_2$ column density is highly uncertain
(e.g., Verter \& Hodge \markcite{VH95}1995; Wilson \markcite{W95}1995).
Thus, we note that the column densities quoted in the following discussion are lower 
limits since we have elected to apply no correction for the contribution of 
molecular gas.  We do, however, include a correction for helium: 
N$_{\rm gas}$ = 4/3 N$_{\rm HI}$.  The observed peak neutral hydrogen gas column densities
(N$_{\rm HI}$) and the corresponding neutral gas surface density (corrected for inclination
and helium content, $\Sigma_{\rm g}$) are tabulated in Table \ref{tab:colden}.

Many of the gas density studies rely on azimuthal averages of the gas surface
density and star--formation rate, such as those illustrated in 
Figure \ref{fig:radprofs}.   In disk systems, the Toomre \markcite{T64}(1964)
instability criterion provides a natural threshold density with which
to compare azimuthally averaged radial density profiles.
The Toomre criterion represents the balance between gravitational collapse
and rotational shear and thermal pressure; we note that in the limit of zero 
rotational shear (solid body rotation), this criterion is still valid. 
We have tabulated the value of the Toomre instability criterion  
($<\Sigma_c>$) for UGC 4483, 
UM 439, and UM 462 in Table \ref{tab:colden}, assuming the kinematic parameters 
from Section \ref{sec:kin} and solid body rotation throughout the disk
(II~Zw~40 and UM 461 have been excluded from this analysis since their
kinematic parameters were unconstrained).  In contrast
to the results for dIs, where the global gas density rarely, if ever, exceeds
the Toomre instability criterion (e.g., van Zee \etal \markcite{vHSB97}1997b; 
 Hunter \etal \markcite{HEB98}1998a), the Toomre value is exceeded in the
central regions of all three BCDs, suggesting that star--formation
activity should be enhanced in the inner regions of these galaxies.  

While radial profiles are potentially useful as global indicators of star--formation
activity, azimuthal averages tend to dampen star formation signatures in systems 
where star formation is a local process, such as irregular and dwarf galaxies
(e.g., van Zee \etal \markcite{vHSB97}1997b). Visual inspection of Figures 
\ref{fig:iizw40}(c)--\ref{fig:um462}(c) indicates
a good correlation between sites of active star formation and local peaks in the
HI column density.  Intensity profiles for both the HI and H$\alpha$ emission
are presented in Figure \ref{fig:cuts}; each profile is centered on the peak
H$\alpha$ emission feature and is oriented such that additional HII regions
are included in the profile. In addition to the five BCDs in this sample, 
a panel for I~Zw~18 is also included (van Zee \etal \markcite{vWHS98}1998). 
Upon detailed inspection of these intensity profiles, it is clear that
the high rate of star--formation activity in the BCDs has had an impact on the 
neutral gas surface density. In particular, while situated near the column density peak,
the brightest HII regions in II Zw 40 and UGC 4483 are actually located in 
slight depressions. This is likely due to either
strong stellar winds evacuating the gas or photoionization from the
UV continuum of the OB stars.   Despite this evidence of  local feedback from 
the star--formation activity on the neutral medium, the overall impression is
that sites of active star formation are associated with regions of high column density
(2--4 $\times 10^{21}$ atoms cm$^{-2}$).   

It has been hypothesized that the empirical critical star--formation threshold
density (10$^{21}$ atoms cm$^{-2}$) found in many types of galaxies
 could correspond to a critical column
density of dust necessary to shield molecular gas from UV radiation 
(Federman \etal \markcite{FGK79}1979).  If so, this suggests that the 
empirical threshold might be a function of metallicity.
The BCD sample which, with the addition of I~Zw~18 (van Zee \etal \markcite{vWHS98}1998),
spans a factor of 10 in metallicity, provides an opportunity to test this
hypothesis.  In Figure \ref{fig:column}(a) the observed peak column densities are
plotted as a function of metallicity for the BCD sample and for dI galaxies from the 
literature (Skillman \etal \markcite{STTv88}1988; Shostak \& Skillman \markcite{SS89}1989; 
Lo \etal \markcite{LSY93}1993; van Zee \etal \markcite{vHSB97}1997b).  The oxygen 
abundances for the dI galaxies and I~Zw~18 were also taken from the literature  
(Skillman \etal \markcite{SKH89}1989; Skillman \& Kennicutt \markcite{SK93}1993;
van Zee \etal \markcite{vHSB96}1996; van Zee \etal \markcite{vHS97}1997a).
While the peak column density may not be the most appropriate indicator (a ``threshold''
density might be more useful, but is rarely reported in the literature), it is a useful 
diagnostic, particularly in BCDs where one can assume that the HII regions have not 
significantly aged.  The peak column densities have not been corrected for inclination or beam
effects, so a large scatter in this plot is not unexpected.  The total number of objects
is still small, but there is no obvious trend for an increase in column density with 
decreasing metallicity; rather, the most obvious trend in this plot is that dIs 
clearly have lower column densities than the BCDs at similar metallicities. 
By removing the restriction of previous knowledge of the oxygen abundance, larger sample
sizes for both the BCDs and dIs are possible.  In Figure \ref{fig:column}(b),
the distributions of peak column densities for BCDs and dIs are shown.  The
BCD sample includes data from the present sample, Taylor \etal \markcite{TBPS94}(1994),
and van Zee \etal \markcite{vWHS98}(1998); the dI sample includes those galaxies
in the above literature compilation as well as  
Skillman \etal \markcite{SBMW87}(1987), C\^ot\'e \markcite{C95}(1995), 
Kobulnicky \& Skillman \markcite{KS95}(1995), and Hunter \etal \markcite{HvG96}(1996).
Again, the observed column densities have
not been corrected for inclination or for beam--smearing effects, but the BCDs,
on average, have higher peak column densities than the dIs.  This trend may
explain, or be a result of, the high rates of star formation in BCDs.

\section{Evolutionary Scenarios}
\label{sec:dwfs}

As recognized by Searle \& Sargent \markcite{SS72}(1972), the
defining characteristic of a BCD -- a bright, compact, star burst --
must be a short lived phenomenon
(see Table \ref{tab:global} for examples of gas depletion timescales).
In fact, the current star--formation rate must be higher than the
average past rate in these systems in order to explain their compact, 
low luminosity, low metallicity nature.  It was thus hypothesized that these
galaxies undergo episodic bursts of star formation, with the corresponding
conclusion that they spend a fraction of their lifetime in a quiescent state.
BCDs in a quiescent state have not yet been found, however, in part due to 
the difficulty of identifying such objects.  For instance, it is still unclear what 
these galaxies should look like since their high rate of star formation will
presumably affect many physical properties such as the ISM, total luminosity, 
colors, etc.   Nonetheless, it is natural to wonder if there may be an evolutionary 
link between the three main classes of low mass galaxies:
dI, dE, and BCDs (e.g., Skillman \markcite{S96}1996).  
 Below we discuss possible evolutionary scenarios
for both pre-- and post--BCDs. 

In general, BCDs and dIs have many similar global properties, 
with the exception of optical surface brightness and optical size 
(e.g., Hoffman \etal \markcite{HHSL89}1989).  For instance,
both dIs and BCDs can have large, extended HI distributions.
 However, in contrast to the extensive
HI envelopes around many quiescent dIs (e.g., van Zee \etal 
\markcite{vHSB97}1997b), the HI envelopes around BCDs have
many kinematic irregularities.  The BCDs may still be
in the process of collapse, or this may be the result
of their high rates of star formation.
Another major difference between BCDs and dIs is that the
BCDs have higher central mass concentrations in both gas
and stellar content.  This latter point has been the major 
stumbling block for simple evolution models between BCDs and dIs
since it indicates that mass redistribution will be
necessary before a BCD can fade into a dI.  
An alternative view might be that BCDs do not evolve into low surface brightness 
dIs at all (because of the differences in their kinematics and gas content), but rather
that their high central mass concentrations simply are normal for
these galaxies.  In this scenario, one would think of dIs as possessing
a range in their central surface brightness/central mass densities, from
very low (e.g., LSB dwarfs) to very high.  The galaxies we see as BCDs
would simply represent the extreme end of this continuum of central
mass densities.  The star burst activity would then be a natural 
consequence of the physical properties of the galaxies: objects with
high central mass concentrations would preferentially be able to
host and sustain a major starburst.

It has also been hypothesized
that BCDs could evolve into dEs, either if the burst uses the
available gas, or if the ISM is disrupted as the burst evolves
(e.g., Dekel \& Silk \markcite{DS86}1986; Davies \& Phillipps \markcite{DP88}1988; 
Marlowe \etal \markcite{MHWS95}1995; Telles \etal \markcite{TMT97}1997).  
The blow--away scenario, however,
is more difficult than previously thought (Mac Low \& Ferrara \markcite{MF98}1998).
Furthermore, while some BCDs are morphologically similar to the dEs, their
kinematics are quite different.  As Ferguson \& Binggeli \markcite{FB94}(1994) 
emphatically state, ``dwarf ellipticals are not supported by rotation.'' 
Admittedly, only a few dEs have been observed, but almost all have
stellar rotation velocities less than 2 \kms~(e.g., NGC 185 and  NGC 205, Bender 
\etal \markcite{BPN91}1991; VCC 351, Bender \& Nieto \markcite{BN90}1990).
The few exceptions (NGC 147, Bender \etal \markcite{BPN91}1991; IC 794,
Bender \& Nieto \markcite{BN90}1990) have rotation velocities of only 10--15 
\kms~which is significantly less than the rotation velocities for the BCDs in
this sample.  Interestingly, the HI gas associated with NGC 205 does have a 
significant velocity gradient, but this gas may have been captured recently 
(Young \& Lo \markcite{YL97}1997).  Nonetheless, it appears that most
dEs are not supported by rotation while many (all?) BCDs are
rotation dominated.  Thus, loss of angular momentum would be necessary
before a BCD could evolve into a dE.  Furthermore, one additional argument
against BCD to dE evolution comes from environmental considerations:
most dEs are found in clusters or as companions to massive galaxies
while BCDs are typically field objects.
In fact, dEs are among the most strongly clustered galaxies known (Ferguson \&
Sandage \markcite{FS89}1989), while BCDs are among the least clustered
(Iovino \etal \markcite{IMS88}1988; Salzer \& Rosenberg \markcite{SR94}1994; 
Pustil'nik \etal \markcite{PULTG95}1995).
If the previous generation of BCDs all evolved into dE galaxies, it
would lead to a large population of unclustered dEs, which is not
observed to exist.  Note that we cannot rule out the possibility that
the present day dEs evolve through a BCD--like phase during their early evolution.  
Rather,  we suggest that the BCDs we see today will not evolve into dEs.

One solution to allow a BCD to dE evolutionary scenario is to
hypothesize that interactions with other galaxies can enable the BCD to lose the
necessary angular momentum.  Interactions have also been invoked to
explain the high rate of star formation in BCDs (e.g., 
Taylor \etal \markcite{TBS93}1993; Taylor \etal \markcite{TBGS95}1995; 
Walter \etal \markcite{WBDK97}1997),
and thus are an extremely appealing mechanism to not only
explain the initiation of star--formation activity, but also the 
subsequent evolution of a BCD.  However, based on a search for optical 
companions, Telles \& Terlevich \markcite{TT95}(1995) found that a 
significant fraction of BCDs were isolated.  Furthermore, of 
the five BCDs in this sample, only one (II~Zw~40) is clearly the result of 
an interaction.  Thus, while an interaction is certainly one mechanism
by which star formation can be enhanced in BCDs, it is  not
the only one, and possibly not even  the most common.

Further observational evidence will be needed to determine the
ultimate fate of BCDs.  For instance, optical surface photometry for 
large, unbiased, samples of dIs would be useful to determine if BCDs
simply form the extreme end of a continuum of dI properties. The study
of Patterson \& Thuan \markcite{PT96}(1996) represents a good start in
this direction, but more data are needed.
  In addition, studies of the gas content
in BCDs, dIs, and dEs would be beneficial for identification of systematic
trends in gas content, and to confirm that most dEs are gas--poor.
Further, detailed studies of the mass distributions (stellar, gas, and 
total) in all types of dwarf galaxies will be necessary to determine 
if BCDs, dIs, and dEs form a common family.  Finally, kinematic
studies, particularly of the stellar distributions in dEs, will be
crucial for understanding the relationships between these fundamental
classes of low mass systems.

\section{Summary and Conclusions}
\label{sec:conc}

In summary, we have presented the results of high spatial resolution 
HI synthesis observations of five BCDs.  Our results and conclusions
are summarized below.

(1)  The HI distributions extend well beyond the optical system, with a
typical D$_{\rm HI}$/D$_{\rm opt}$ of 3--4.  However, based on the
velocity information, some of the extended gas is not in dynamical
equilibrium.  In fact, whereas velocity gradients were detected in all
five objects, none are smoothly rotating systems.

(2) The HI distributions are centrally peaked, in contrast to those
of dIs (e.g., van Zee \etal \markcite{vHSB97}1997b).  In all cases
the central column density exceeds the Toomre \markcite{T64}(1964)
instability criterion.

(3) Detailed analysis of the HI distributions indicates that the neutral
gas is clumpy on spatial scales of hundreds of parsec.  As seen previously
in other dwarf galaxy systems, the peak gas column densities are associated
with sites of active star formation.  Furthermore,  moderate local depressions 
in the neutral gas column density are associated with the most intense
knots of star formation; in these regions, the column density drops 20--50\%,
presumably due to feedback from the star forming region.
 
(4) The observed neutral gas velocity dispersions, $\sim$11 \kms,
are not unusual for star forming dwarf galaxies.  With typical rotation velocities
of 20--40 \kms, these systems are {\it rotation dominated}.
 
These observations have shed new light
on the evolutionary connections between BCDs, dIs, and dEs.  In particular,
BCDs appear to have intrinsically different neutral gas distributions
than the more quiescent dIs, perhaps explaining their higher rates of star formation. 
Furthermore, if BCDs are rotationally supported, as was found here for all five
systems, BCDs cannot evolve into dEs without substantial loss of angular
momentum.   Thus, the neutral gas kinematics and distribution place strong
constraints on the feasibility of many of the popular BCD evolutionary scenarios.

\acknowledgements
We thank Chris Taylor, Elias Brinks and Lisa Young for comments on a preliminary
version of this paper. We also thank our anonymous referee for suggestions
which improved the presentation of the paper.  We thank Chris Taylor for providing 
us with  data from previous BCD studies.  We acknowledge the financial 
support by NSF grant AST95--53020 to JJS.  Partial support from
a NASA LTSARP grant No. NAGW--3189 (EDS) is gratefully acknowledged.
This research has made use of the NASA/IPAC Extragalactic Database (NED)   
which is operated by the Jet Propulsion Laboratory, California Institute   
of Technology, under contract with the National Aeronautics and Space      
Administration.

\vfill
\eject
\psfig{figure=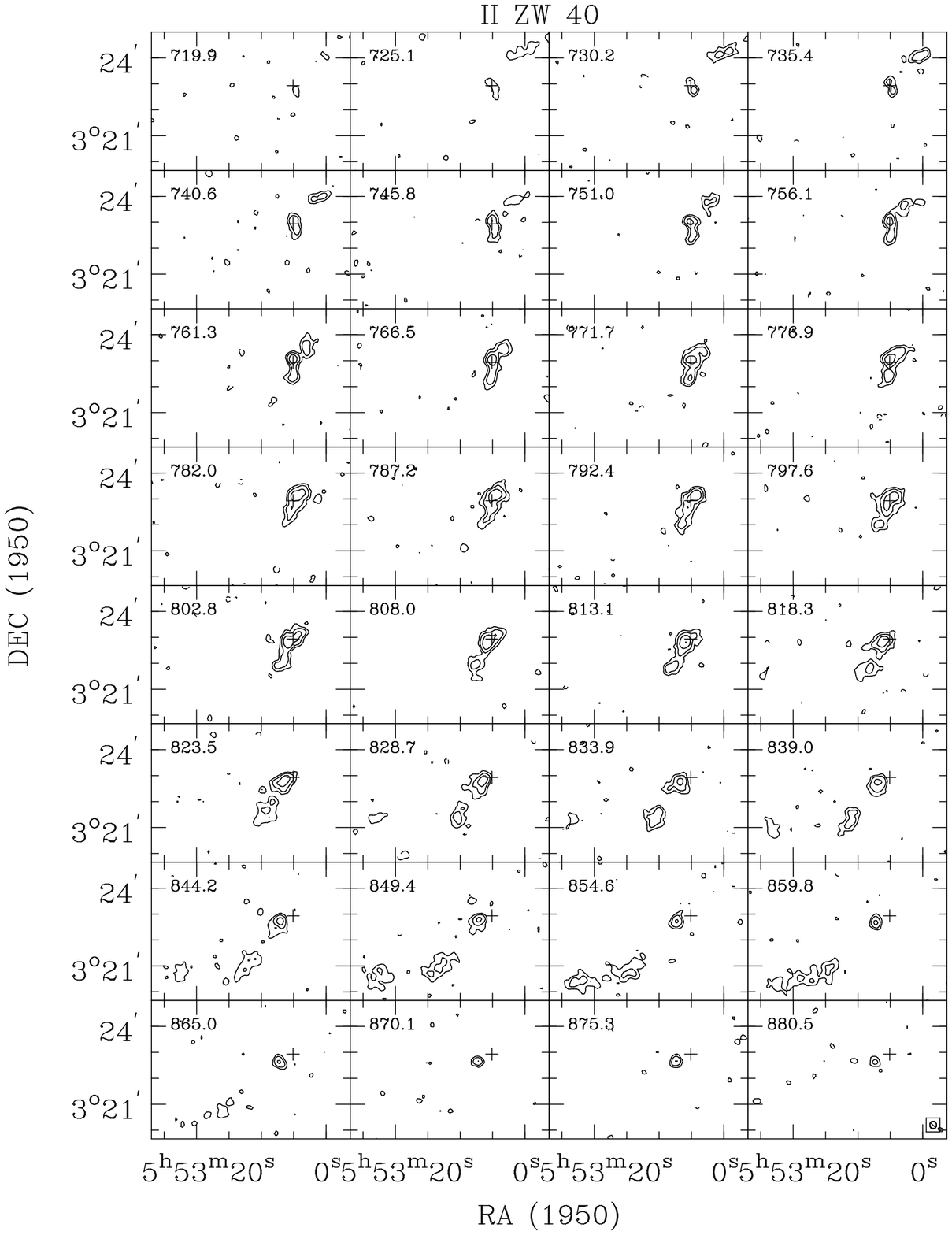,width=5.7in,bbllx=100pt,bblly=50pt,bburx=600pt,bbury=670pt,clip=t}
\vskip -0.2 truein
\figcaption[vanzee.fig01.ps] { Selected channels from the tapered data cube of II~Zw~40.  
Every other channel is shown. The beam size, illustrated in the lower right panel,
 is 17.0$\times$15.4 arcsec.
The contours represent --3$\sigma$, 3$\sigma$, 
6$\sigma$, and 12$\sigma$.  The cross indicates the location of the nominal
kinematic center. \label{fig:chans40} }

\psfig{figure=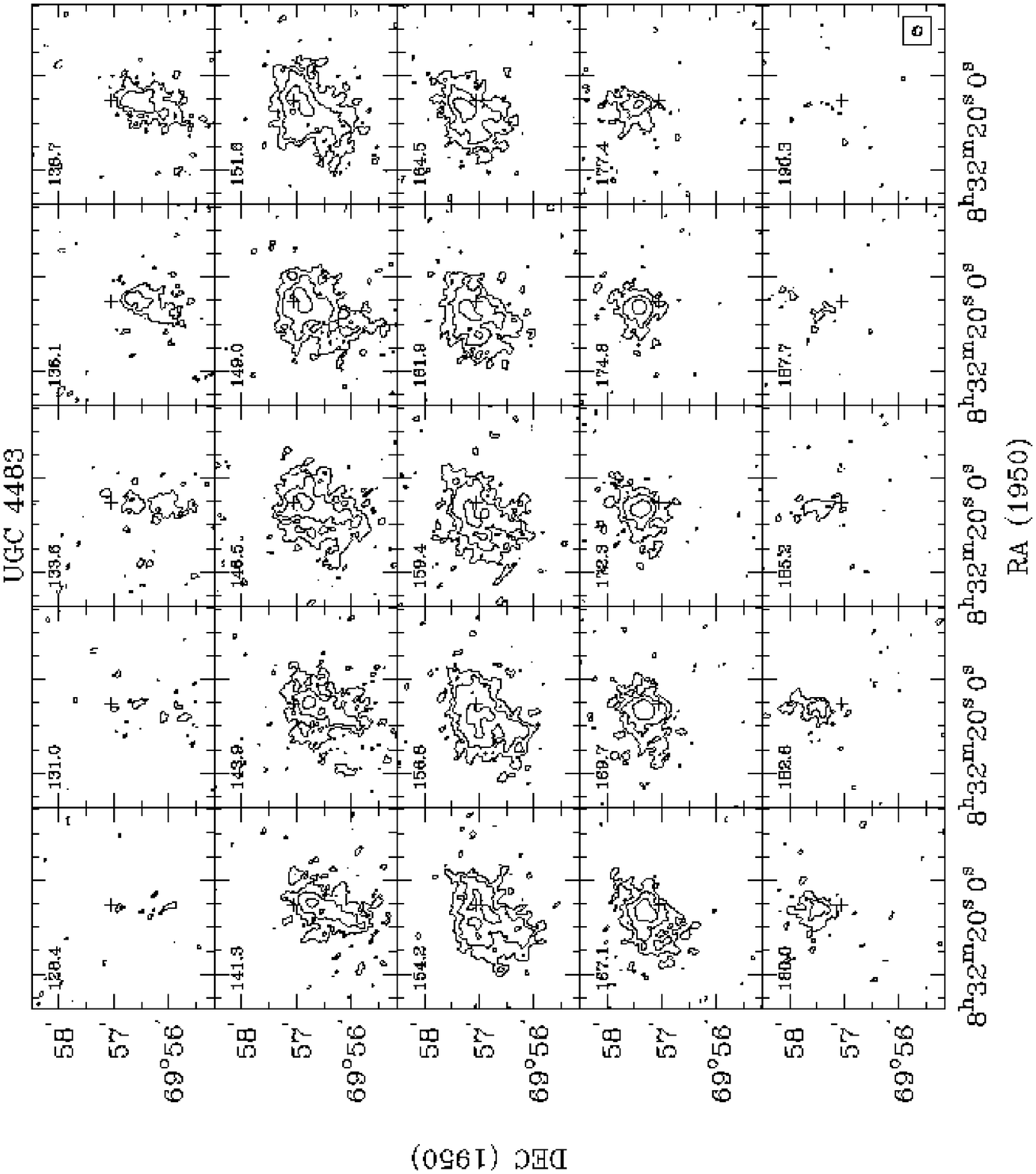,width=7.5in,angle=-90.}
\figcaption[vanzee.fig02.ps] { Selected channels from the natural weight data cube of UGC 4483.  
The beam size, illustrated in the lower right panel, is 10.8$\times$8.9 arcsec.
The contours represent --3$\sigma$, 3$\sigma$, 
6$\sigma$, and 12$\sigma$.  The cross indicates the location of the nominal
kinematic center. \label{fig:chans4483} }

\psfig{figure=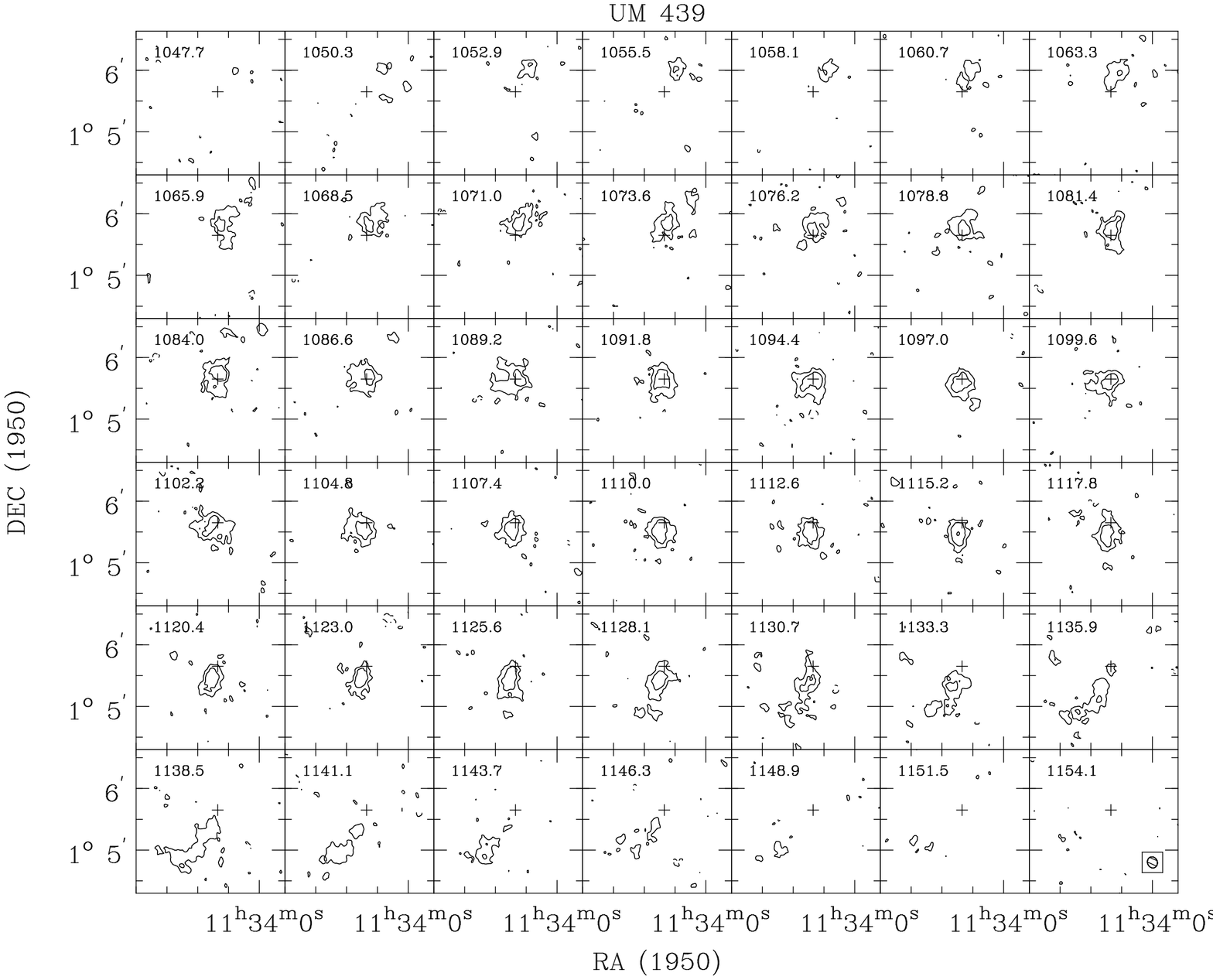,width=7.in,angle=-90.}
\figcaption[vanzee.fig03.ps] { Selected channels from the natural weight data cube of UM 439.  
The beam size, illustrated in the lower right panel, is 11.0$\times$9.6 arcsec.
The contours represent --3$\sigma$, 3$\sigma$, and
6$\sigma$.  The cross indicates the location of the nominal
kinematic center. \label{fig:chans439} }

\psfig{figure=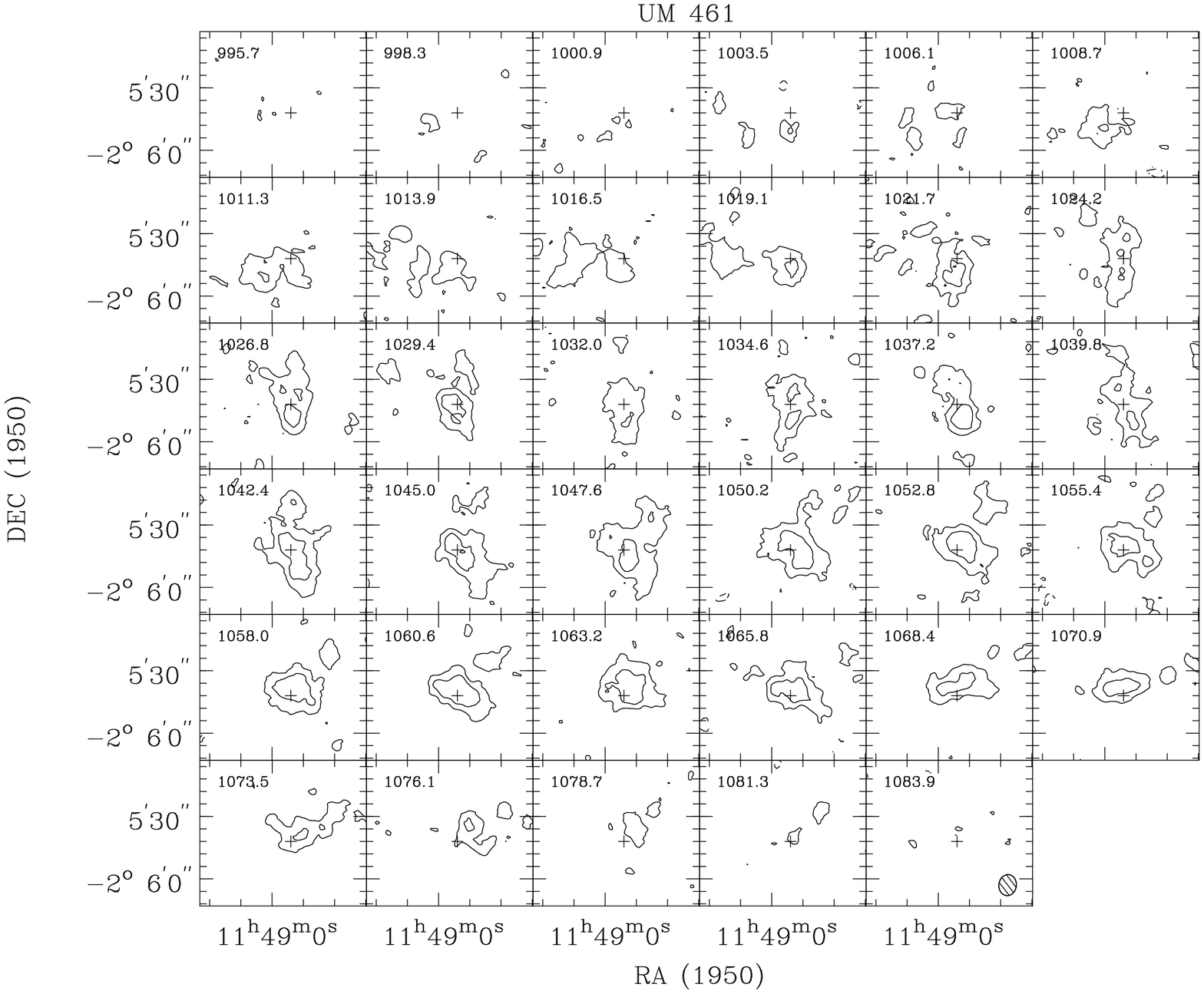,width=7.in,angle=-90.}
\figcaption[vanzee.fig04.ps] { Selected channels from the natural weight data cube of UM 461.  
The beam size, illustrated in the lower right panel, is 10.1$\times$8.4 arcsec.
The contours represent --3$\sigma$, 3$\sigma$, and 
6$\sigma$.  The cross indicates the location of the nominal
kinematic center. \label{fig:chans461} }

\psfig{figure=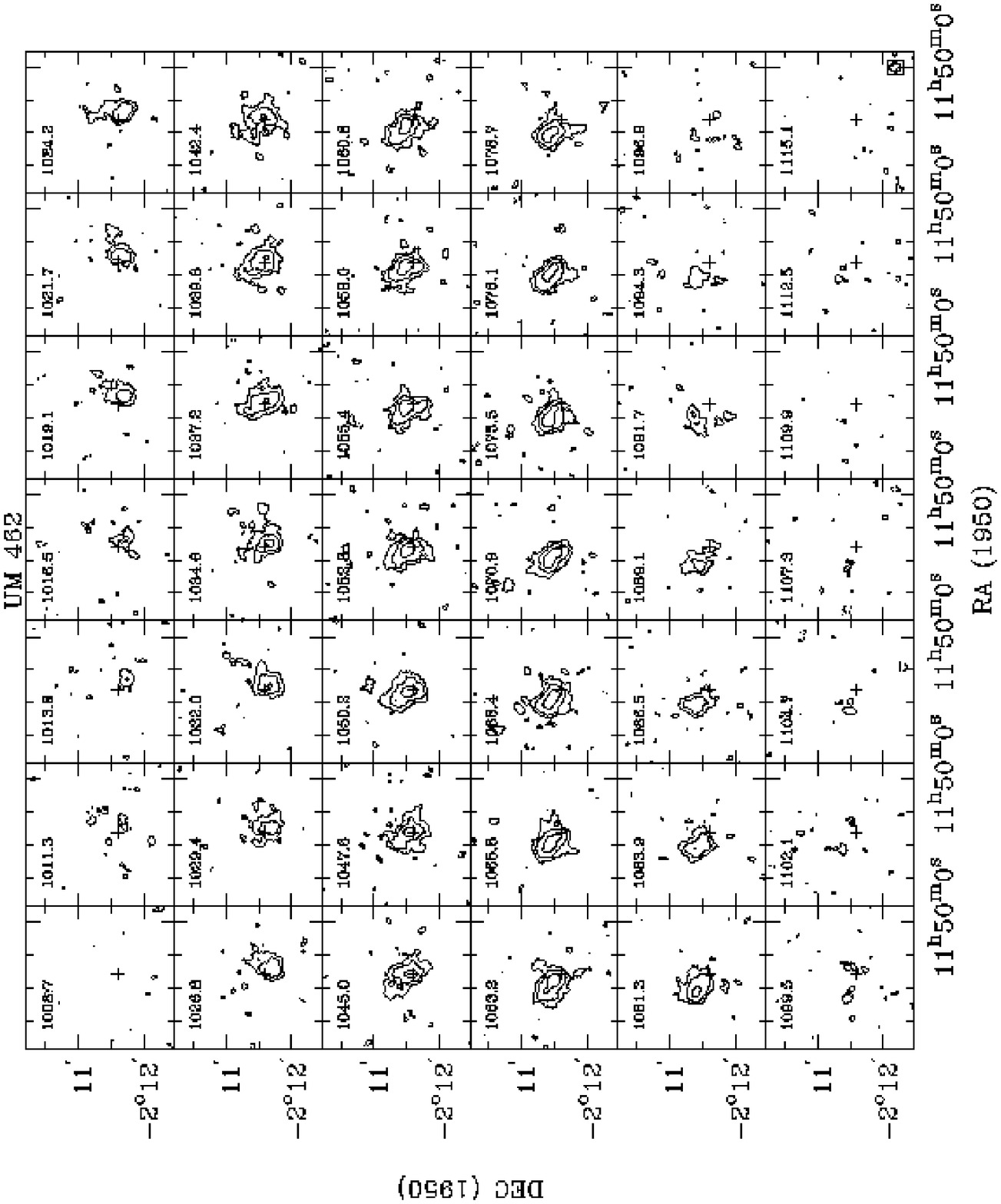,width=6.2in,angle=-90.}
\figcaption[vanzee.fig05.ps] { Selected channels from the natural weight data cube of UM 462.  
The beam size, illustrated in the lower right panel, is 10.1$\times$8.4 arcsec.
The contours represent --3$\sigma$, 3$\sigma$, 
6$\sigma$, and 12$\sigma$.  The cross indicates the location of the nominal
kinematic center. \label{fig:chans462} }

\psfig{figure=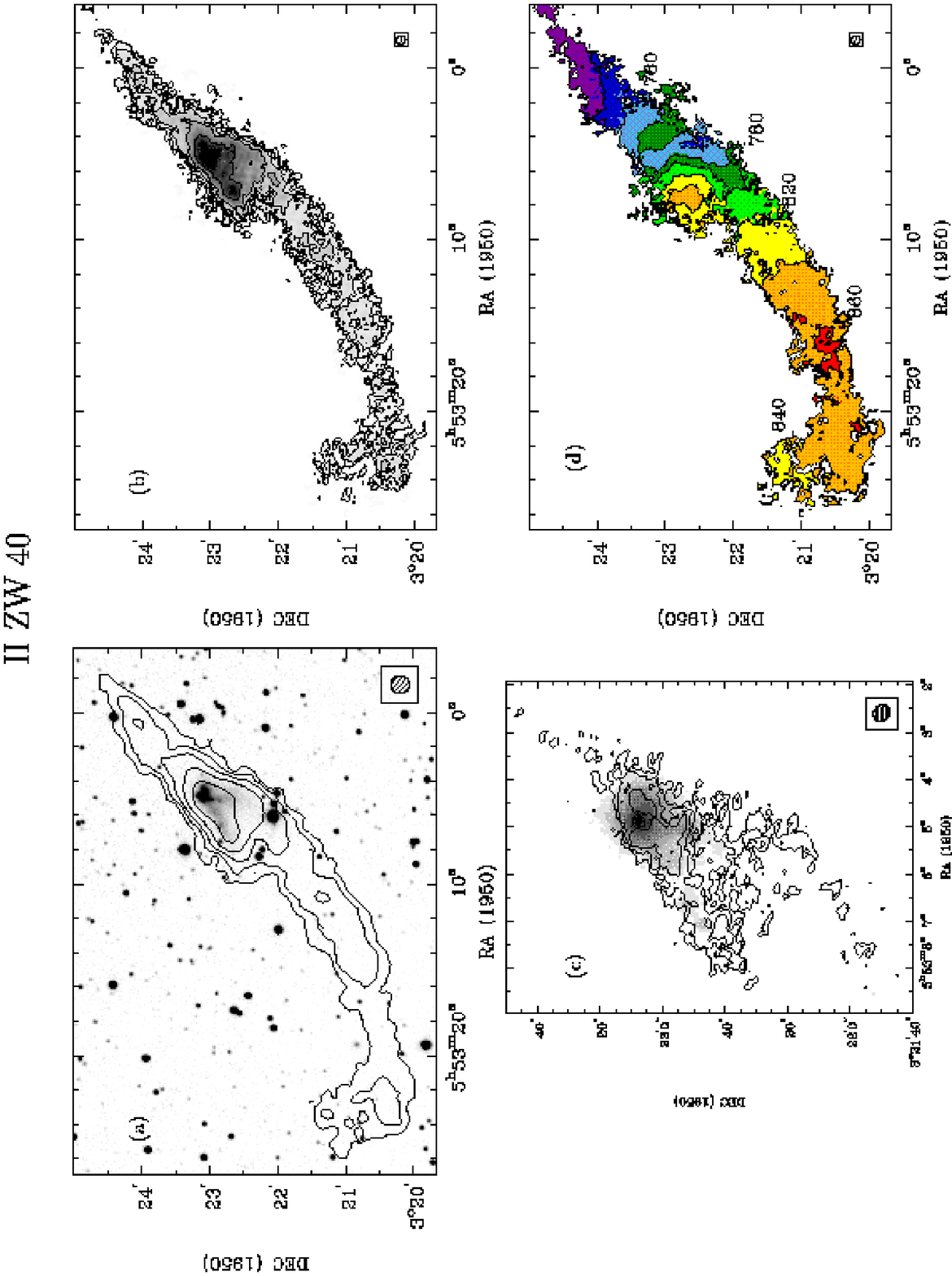,width=6.in,bbllx=1pt,bblly=1pt,bburx=630pt,bbury=900pt,clip=t}
\vfill
\eject
\figcaption[vanzee.fig06.ps]{ Moment maps of II~Zw~40.
(a) The HI contours from the tapered data cube are shown
overlayed on an R band image.  The HI contours
are 1.6, 3.2, 6.4, 12.8, and 25.6 $\times$ 10$^{20}$ atoms cm$^{-2}$ with
a  beam size of 17.0$\times$15.4 arcsec.  The pixel scale of the 
optical image is 0.688 arcsec pixel$^{-1}$.  
(b) The integrated intensity map from the natural weight data cube.
The HI contours are 2, 4, 8, 16, and 32 $\times$ 10$^{20}$ atoms cm$^{-2}$.
The beam size is 8.6$\times$7.6 arcsec.
(c) The HI contours from the intermediate weight data cube overlayed on
an H$\alpha$ image.  The HI contours are 1, 2, and 4 $\times$ 10$^{21}$ atoms 
cm$^{-2}$ with a beam size of 5.7$\times$4.8.  The H$\alpha$ image is
displayed on a logarithmic scale; the pixel scale is 0.43 arcsec pixel$^{-1}$.
(d) The velocity field from the natural weight data cube.  
The contours are marked every 20 \kms.
\label{fig:iizw40} }
\vfill
\eject

\psfig{figure=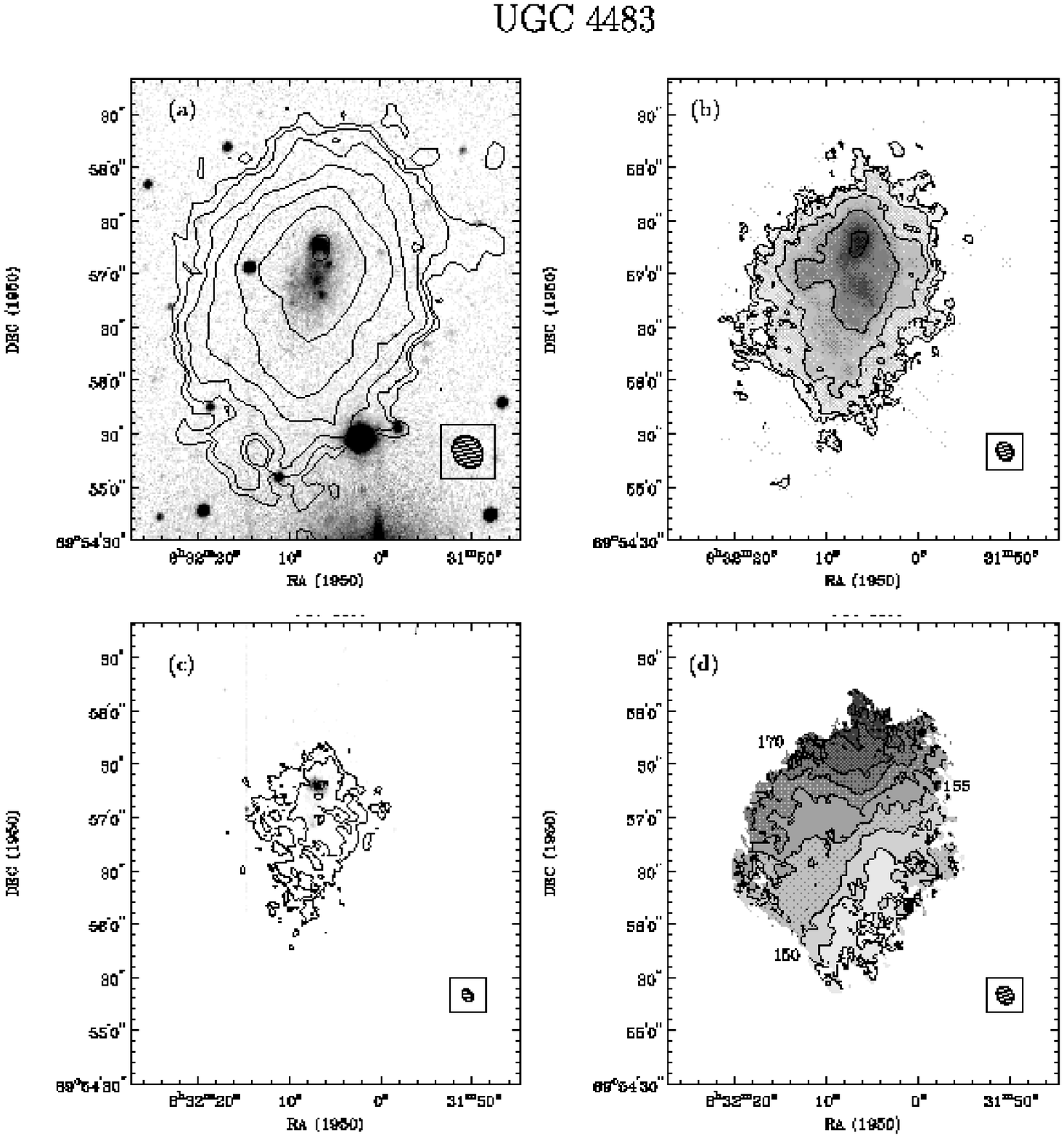,width=6.in,bbllx=1pt,bblly=10pt,bburx=620pt,bbury=750pt,clip=t}
\vskip -0.8truein
\figcaption[vanzee.fig07.ps]{ Moment maps of UGC 4483.
(a) The HI contours from the tapered data cube are shown
overlayed on an R band image.
The HI contours are 4, 8, 16, 32, 64, 128, and 256 $\times$ 10$^{19}$ atoms cm$^{-2}$
with a beam size of  19.3$\times$16.1 arcsec.  The pixel scale of the 
optical image is 0.688 arcsec pixel$^{-1}$.  
(b) The integrated intensity map from the natural weight data cube.
The HI contours are 2, 4, 8, 16, and 32 $\times$ 10$^{20}$ atoms cm$^{-2}$.
The beam size is 10.8$\times$8.9 arcsec.
(c)  The HI contours from the intermediate weight data cube overlayed on
an H$\alpha$ image.  The HI contours are 1, 2, and 4 $\times$ 10$^{21}$ atoms cm$^{-2}$
with a beam size of 7.3$\times$5.3.  The pixel scale of the H$\alpha$ image
is 0.30 arcsec pixel$^{-1}$.
(d) The velocity field from the natural weight data cube.  The contours are marked 
every 5 \kms.
\label{fig:u4483} }

\psfig{figure=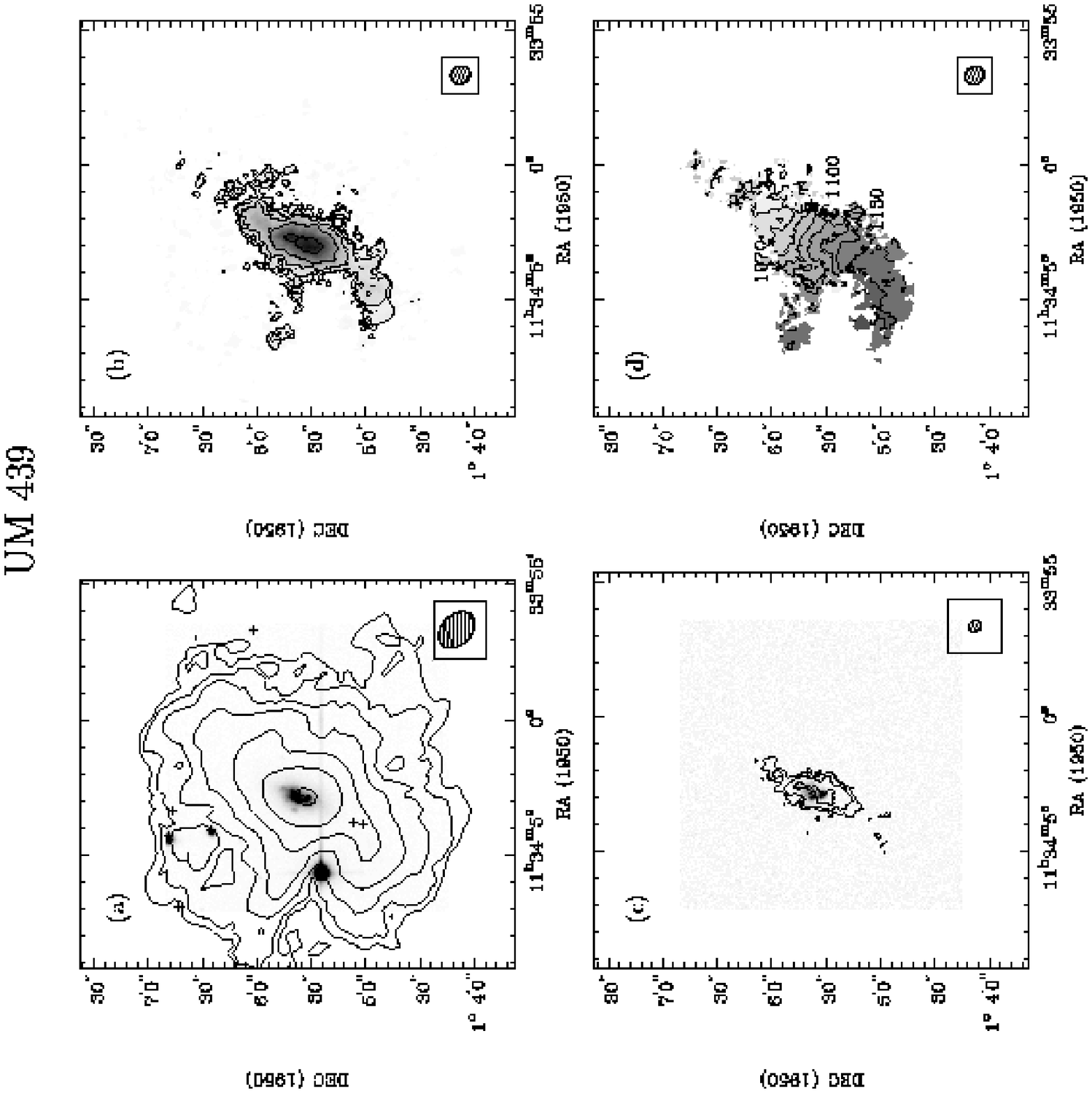,width=6.in,bbllx=1pt,bblly=10pt,bburx=600pt,bbury=700pt,clip=t}
\vfill
\eject
\figcaption[vanzee.fig08.ps]{ Moment maps of UM 439.
(a) The HI contours from the tapered data cube are shown
overlayed on an B band image.
The HI contours are 4, 8, 16, 32, 64, 128, and 256 $\times$ 10$^{19}$ atoms cm$^{-2}$
with a beam size of 23.3$\times$16.3 arcsec.  The pixel scale of the 
optical image is 0.43 arcsec pixel$^{-1}$.  
(b) The integrated intensity map from the natural weight data cube.
The HI contours are 2, 4, 8, 16, and 32 $\times$ 10$^{20}$ atoms cm$^{-2}$.
The beam size is 11.0$\times$9.6 arcsec.
(c)  The HI contours from the intermediate weight data cube overlayed on
an H$\alpha$ image.  The HI contours are 1, 2, and 4 $\times$ 10$^{21}$ atoms cm$^{-2}$
with a beam size of 7.2$\times$5.8.  The pixel scale of the H$\alpha$ image
is 0.43 arcsec pixel$^{-1}$.
 (d) The velocity field of the natural weight data cube.  The contours are marked 
every 10 \kms.
\label{fig:um439} }

\psfig{figure=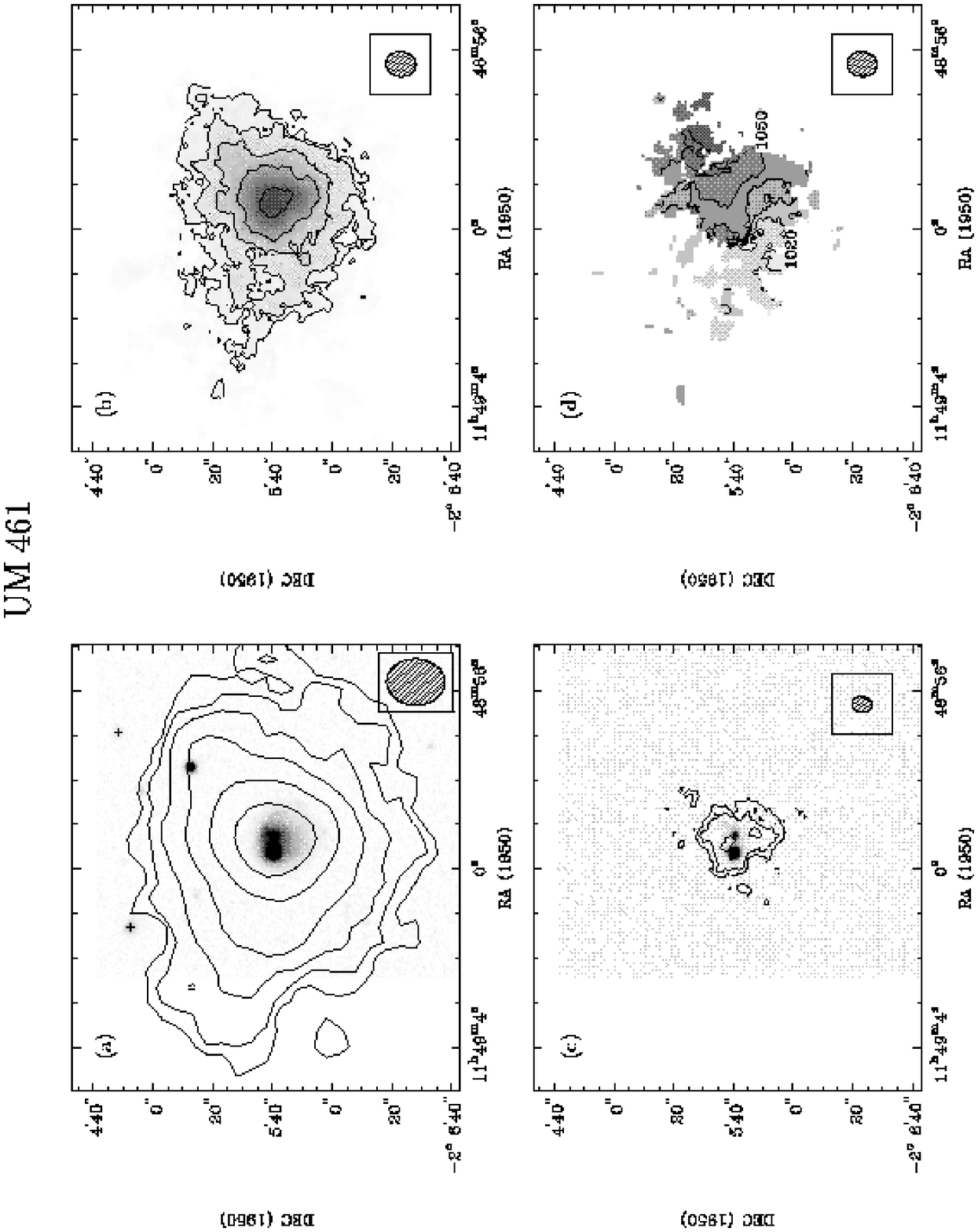,width=6.in,bbllx=1pt,bblly=100pt,bburx=600pt,bbury=1000pt,clip=t}
\vfill
\eject
\figcaption[vanzee.fig09.ps]{ Moment maps of UM 461.
(a) The HI contours from the smoothed natural weight data cube are shown
overlayed on an B band image.
The HI contours are 4, 8, 16, 32, 64, 128, and 256 $\times$ 10$^{19}$ atoms cm$^{-2}$
with a beam size of 19.6$\times$15.5 arcsec.  The pixel scale of the 
optical image is 0.43 arcsec pixel$^{-1}$.  
(b) The integrated intensity map from the natural weight data cube.
The HI contours are 2, 4, 8, 16, and 32 $\times$ 10$^{20}$ atoms cm$^{-2}$.
The beam size is 10.1$\times$8.4 arcsec.
(c) The HI contours from the intermediate weight data cube overlayed on
an H$\alpha$ image.  The HI contours are 1, 2, and 4 $\times$ 10$^{21}$ atoms cm$^{-2}$ 
with a beam size of 6.6$\times$5.2.  The pixel scale of the H$\alpha$ image
is 0.43 arcsec pixel$^{-1}$.
(d) The velocity field of the natural weight data cube.  The contours are marked 
every 10 \kms.
\label{fig:um461} }

\psfig{figure=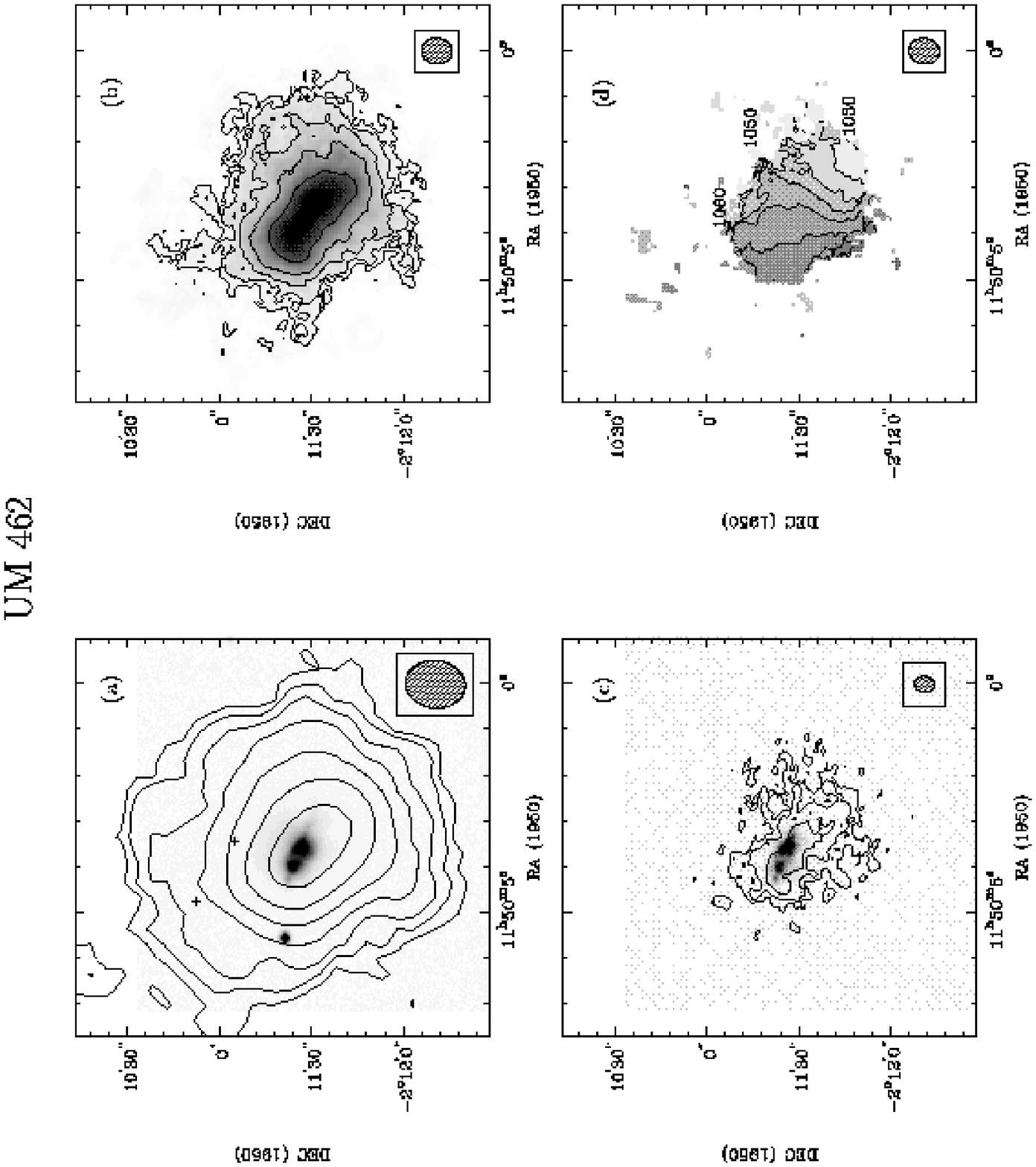,width=6.in,bbllx=1pt,bblly=100pt,bburx=600pt,bbury=900pt,clip=t}
\vfill
\eject
\figcaption[vanzee.fig10.ps]{ Moment maps of UM 462.
(a) The HI contours from the tapered data cube are shown
overlayed on an B band image.
The HI contours are 4, 8, 16, 32, 64, 128, and 256 $\times$ 10$^{19}$ atoms cm$^{-2}$
with a beam size of 19.6$\times$15.5 arcsec.  The pixel scale of the 
optical image is 0.43 arcsec pixel$^{-1}$.  
(b) The integrated intensity map from the natural weight data cube.
The HI contours are 2, 4, 8, 16, and 32 $\times$ 10$^{20}$ atoms cm$^{-2}$.
The beam size is 10.1$\times$8.4 arcsec.
(c) The HI contours from the intermediate weight data cube overlayed on
an H$\alpha$ image.  The HI contours are 1, 2, and 4 $\times$ 10$^{21}$ atoms cm$^{-2}$
with a beam size of 6.6$\times$5.2.  The pixel scale of the H$\alpha$ image
is 0.43 arcsec pixel$^{-1}$.
 (d) The velocity field of the natural weight data cube.  The contours are marked 
every 10 \kms.
\label{fig:um462} }

\psfig{figure=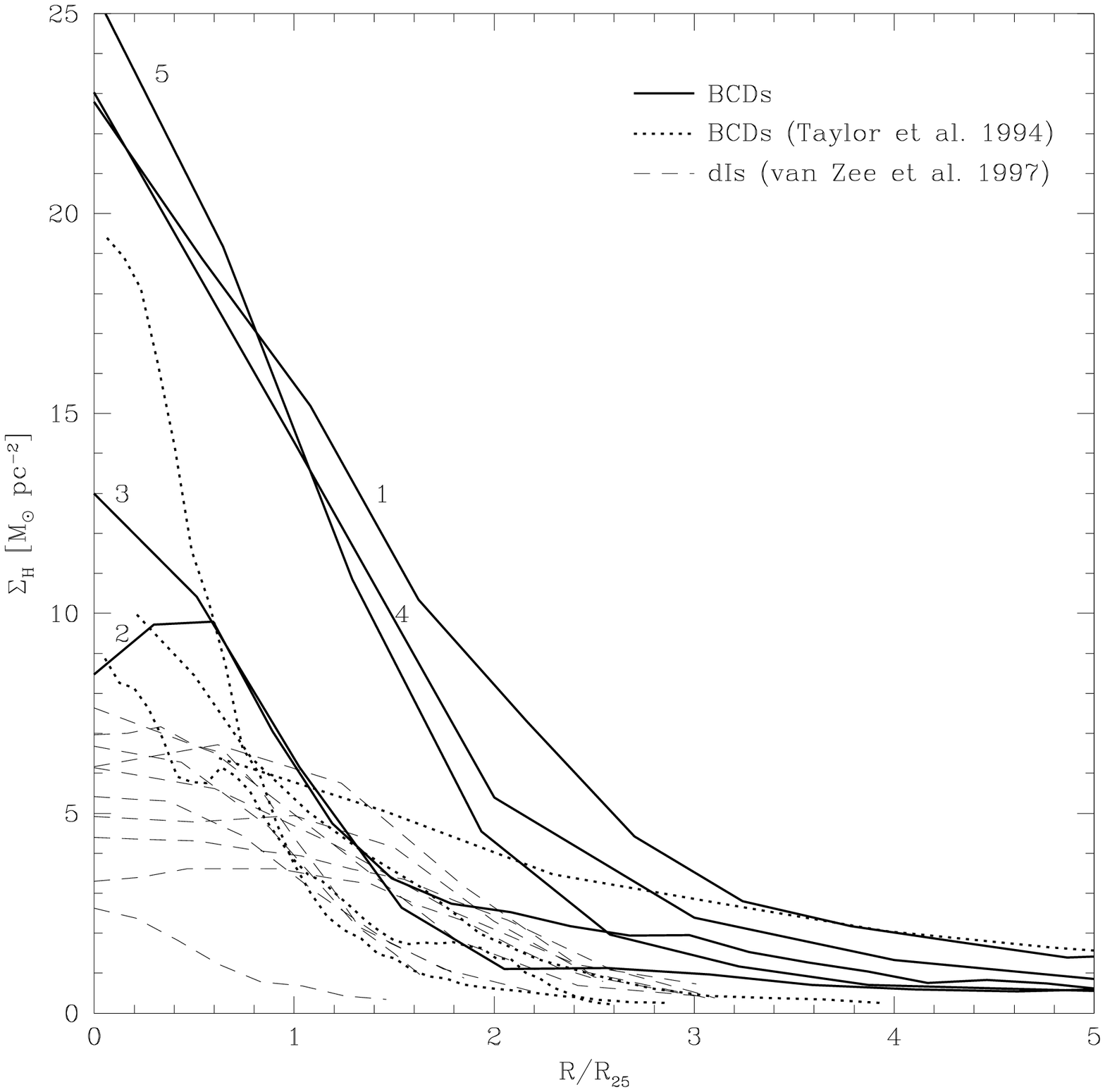,width=6.in,bbllx=1pt,bblly=10pt,bburx=600pt,bbury=700pt,clip=t}
\vskip -1.5 truein
\figcaption[vanzee.fig11.ps]{Radial HI surface densities for BCDs and dIs.  The BCDs from this paper
are denoted by the solid lines: 1-- II Zw 40; 2-- UGC 4483; 3-- UM 439; 4-- UM 461; 
5-- UM 462.
The BCDs from the sample of Taylor \etal (1994) are denoted by dotted lines.  The
dIs from the sample of van Zee \etal (1997b) are denoted by long dashed lines.  
 \label{fig:radprofs} }

\centerline{\hbox{
\psfig{figure=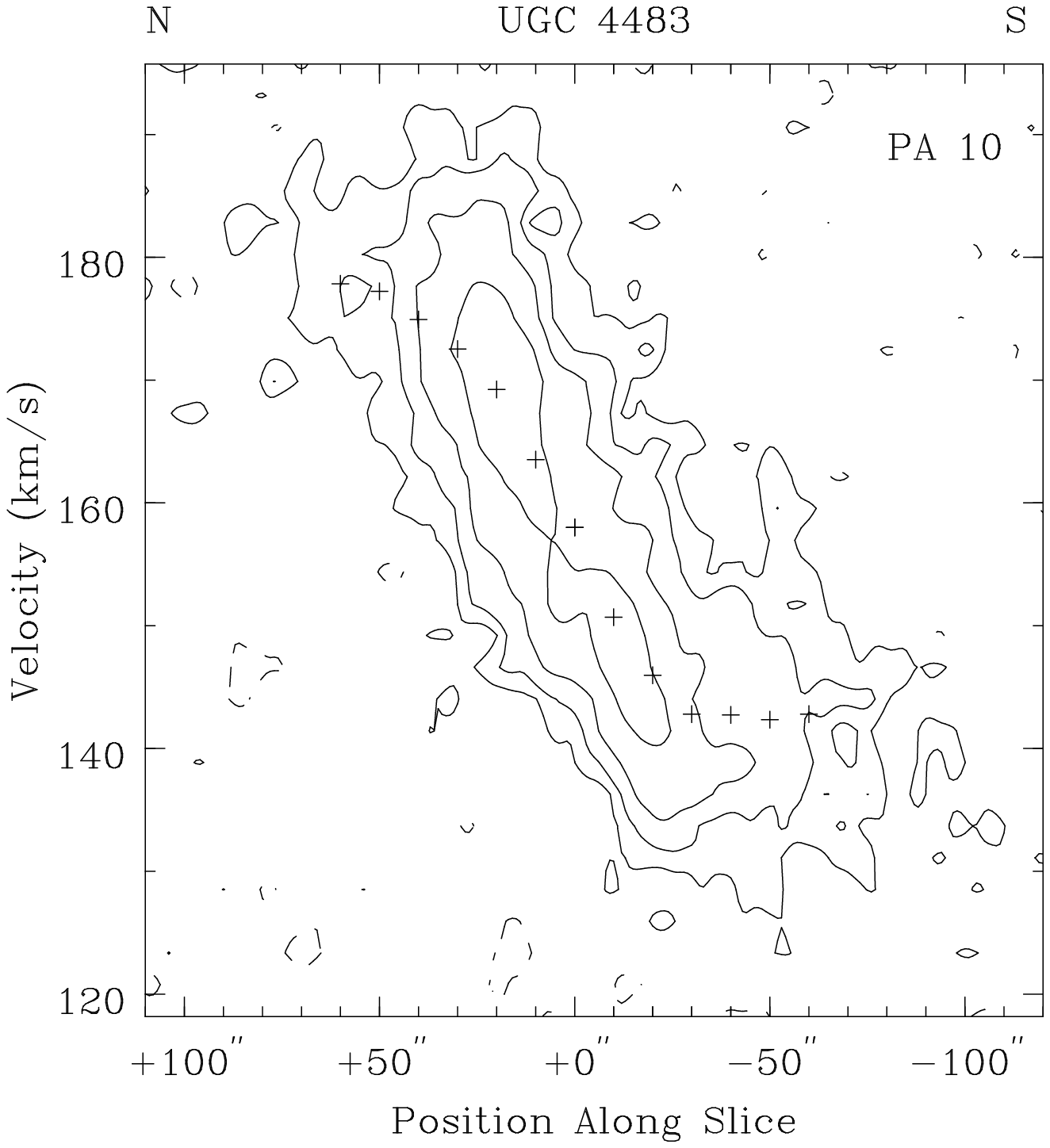,height=3.9in,bbllx=80pt,bblly=20pt,bburx=550 pt,bbury=490 pt,clip=t}
\psfig{figure=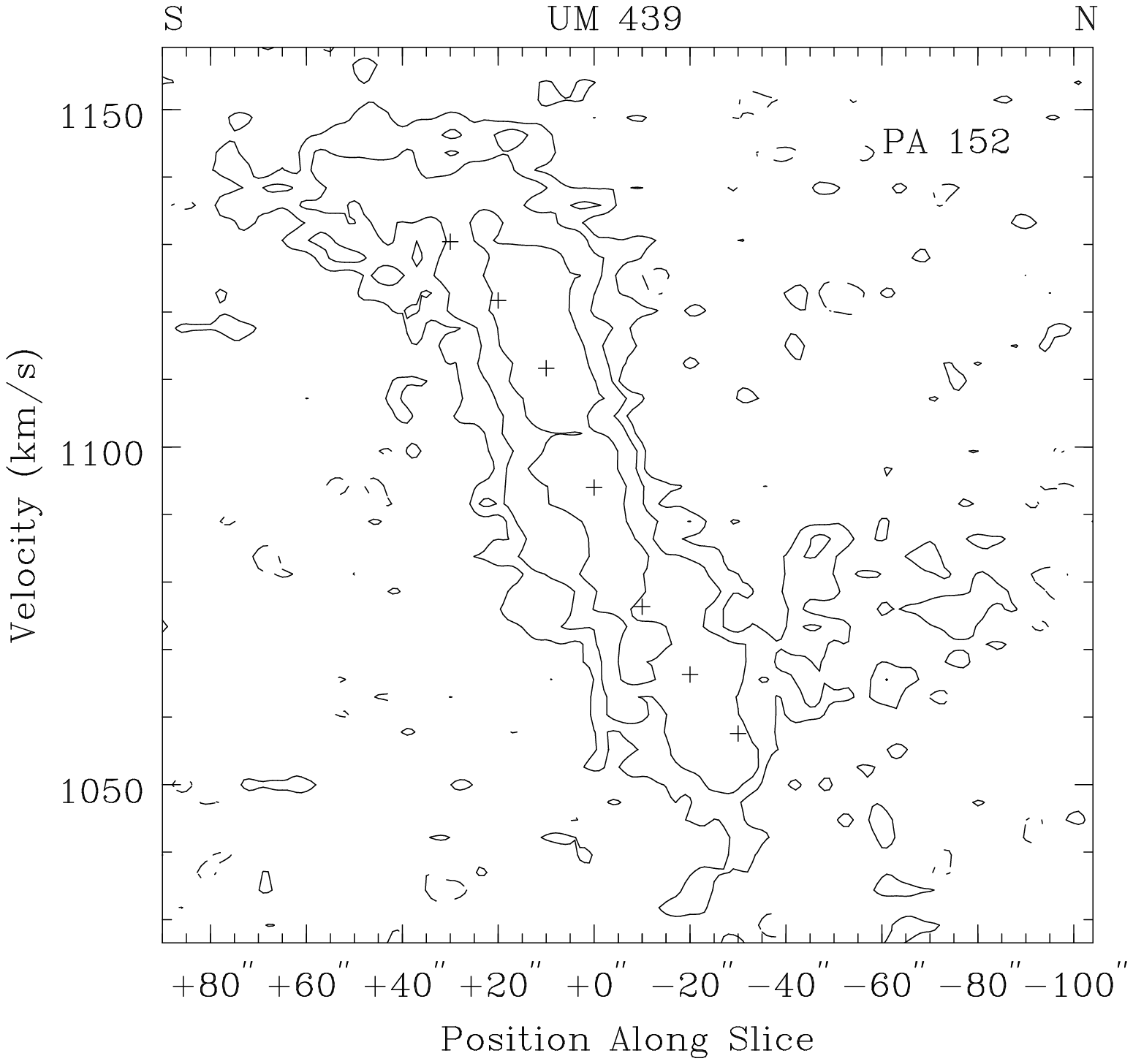,height=3.9in,bbllx=80pt,bblly=20pt,bburx=550 pt,bbury=490 pt,clip=t}
}}

\centerline{\hbox{
\psfig{figure=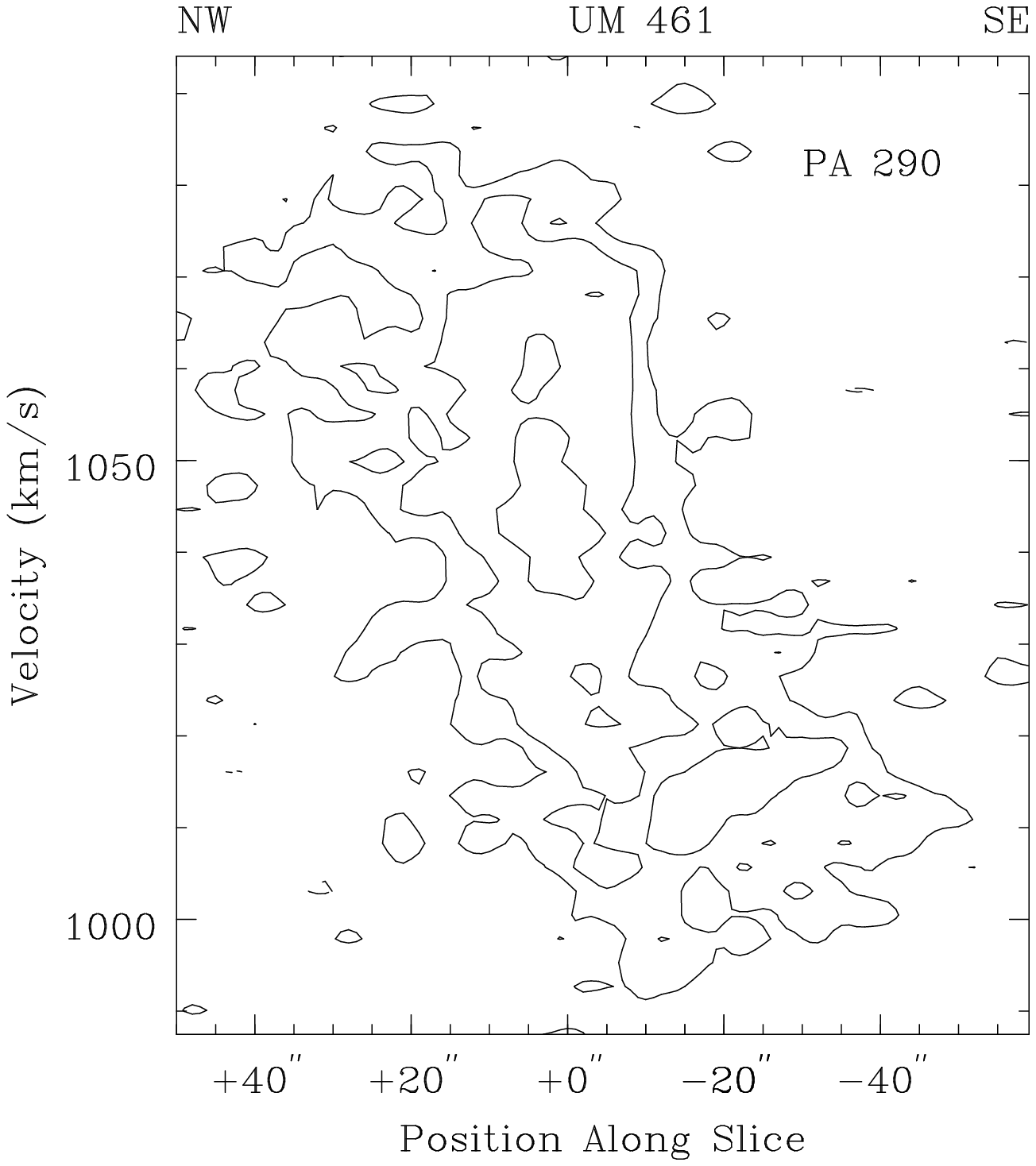,height=3.9in,bbllx=80pt,bblly=20pt,bburx=550 pt,bbury=490 pt,clip=t}
\psfig{figure=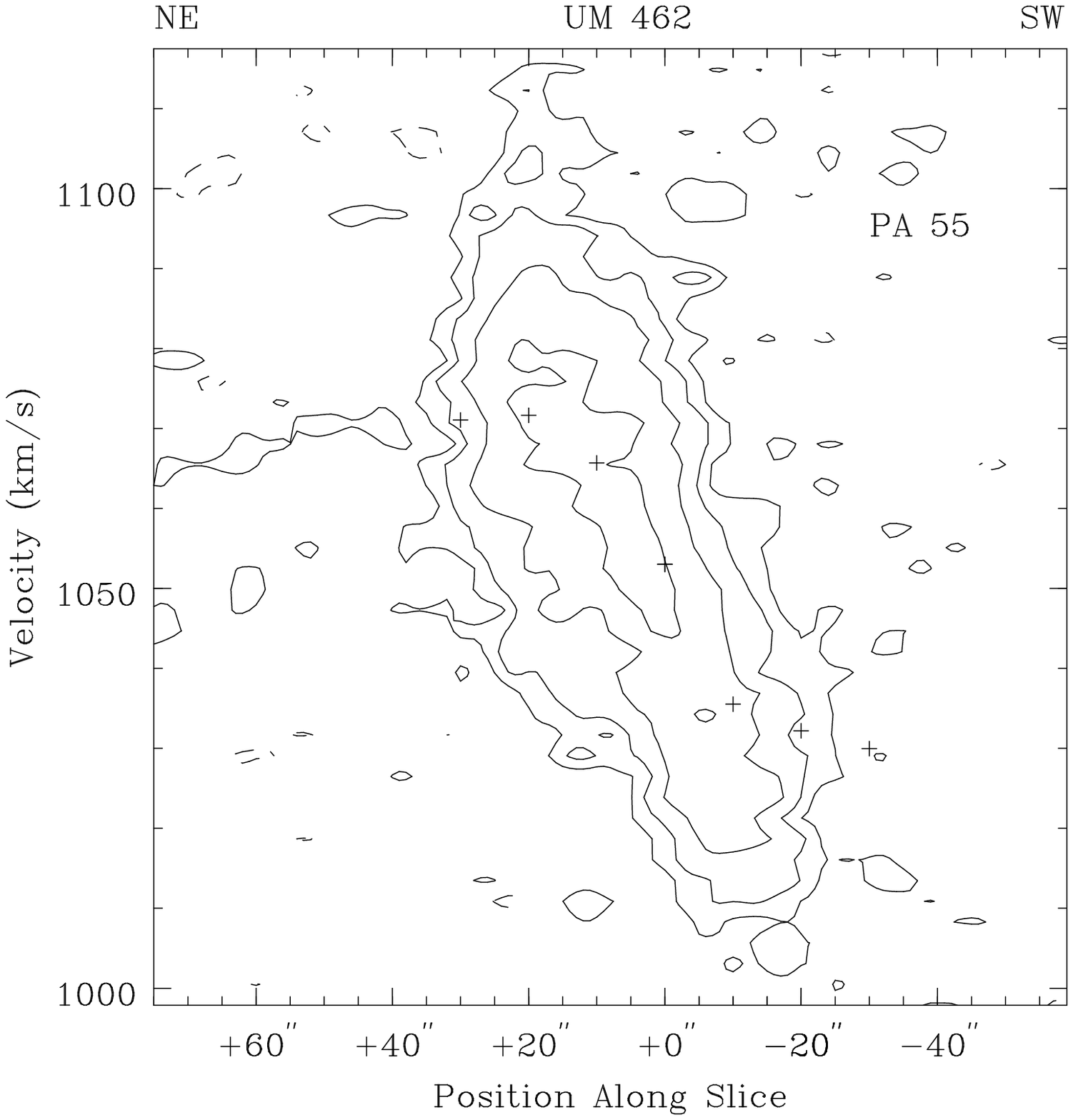,height=3.9in,bbllx=80pt,bblly=20pt,bburx=560 pt,bbury=540 pt,clip=t}
}}
\vskip -7.45truein
\hskip -0.1truein  (a)
\hskip 3.85truein  (b)
\vskip 3.7truein  
\hskip 0.1truein  (c)
\hskip 3.6truein  (d)
\vskip 2.9 truein
\figcaption[vanzee.fig12.ps] {Position--Velocity diagrams from the natural weight data cubes.
The contours represent -3$\sigma$, 3$\sigma$, 6$\sigma$, 12$\sigma$, and 24$\sigma$.
  For all except UM 461, the rotation curves derived from tilted--ring 
models are superposed.\label{fig:pv}}

\psfig{figure=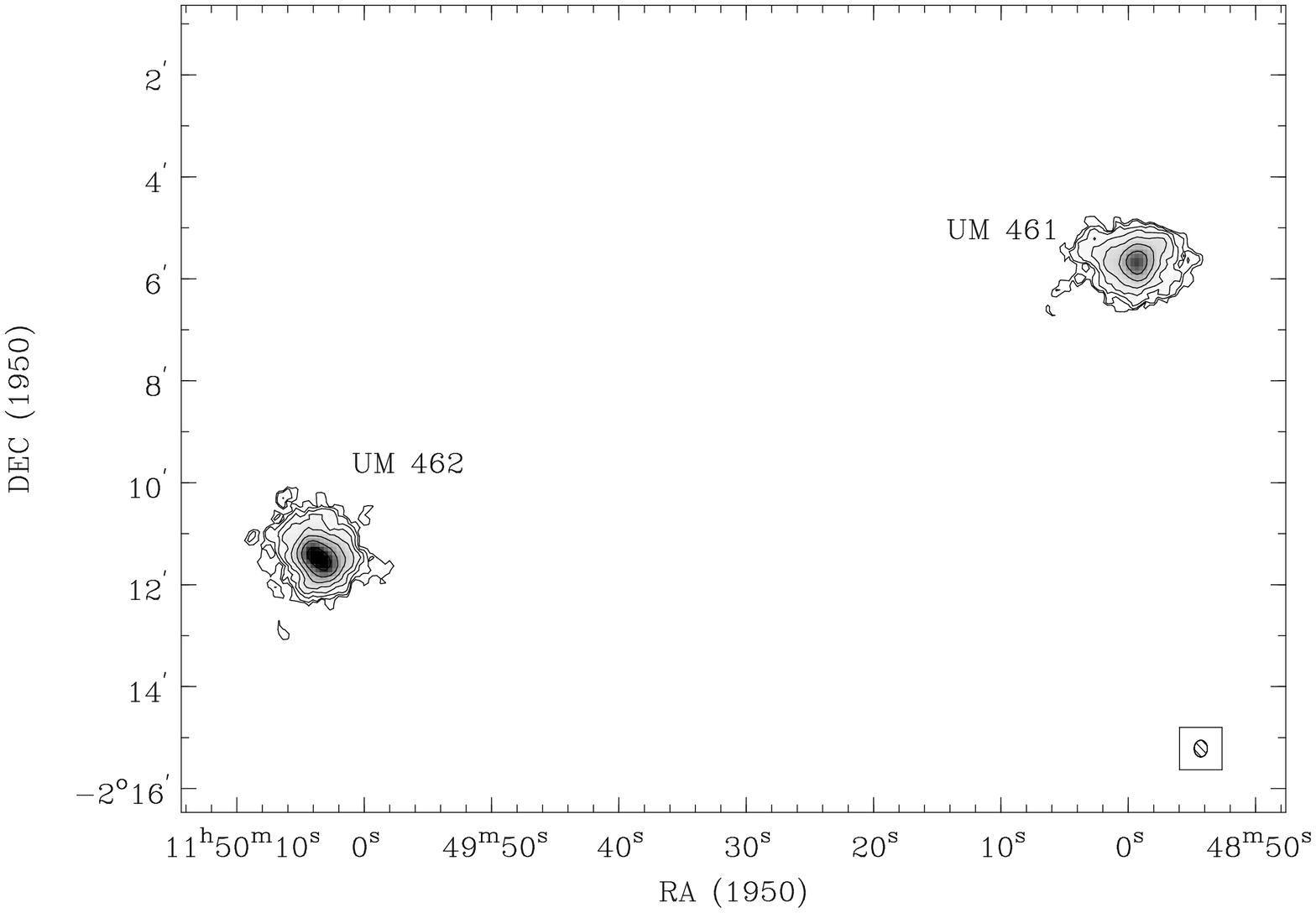,width=7.in,angle=-90.}
\figcaption[vanzee.fig13.ps]{ Moment map of the UM 461/2 system from the tapered data cube.
The HI contours are 2, 4, 8, 16, 32, 64, 128, and 256 $\times$ 10$^{19}$ atoms cm$^{-2}$.  
The beam size of the HI data is 19.6$\times$15.5 arcsec.  \label{fig:um461/2} }

\psfig{figure=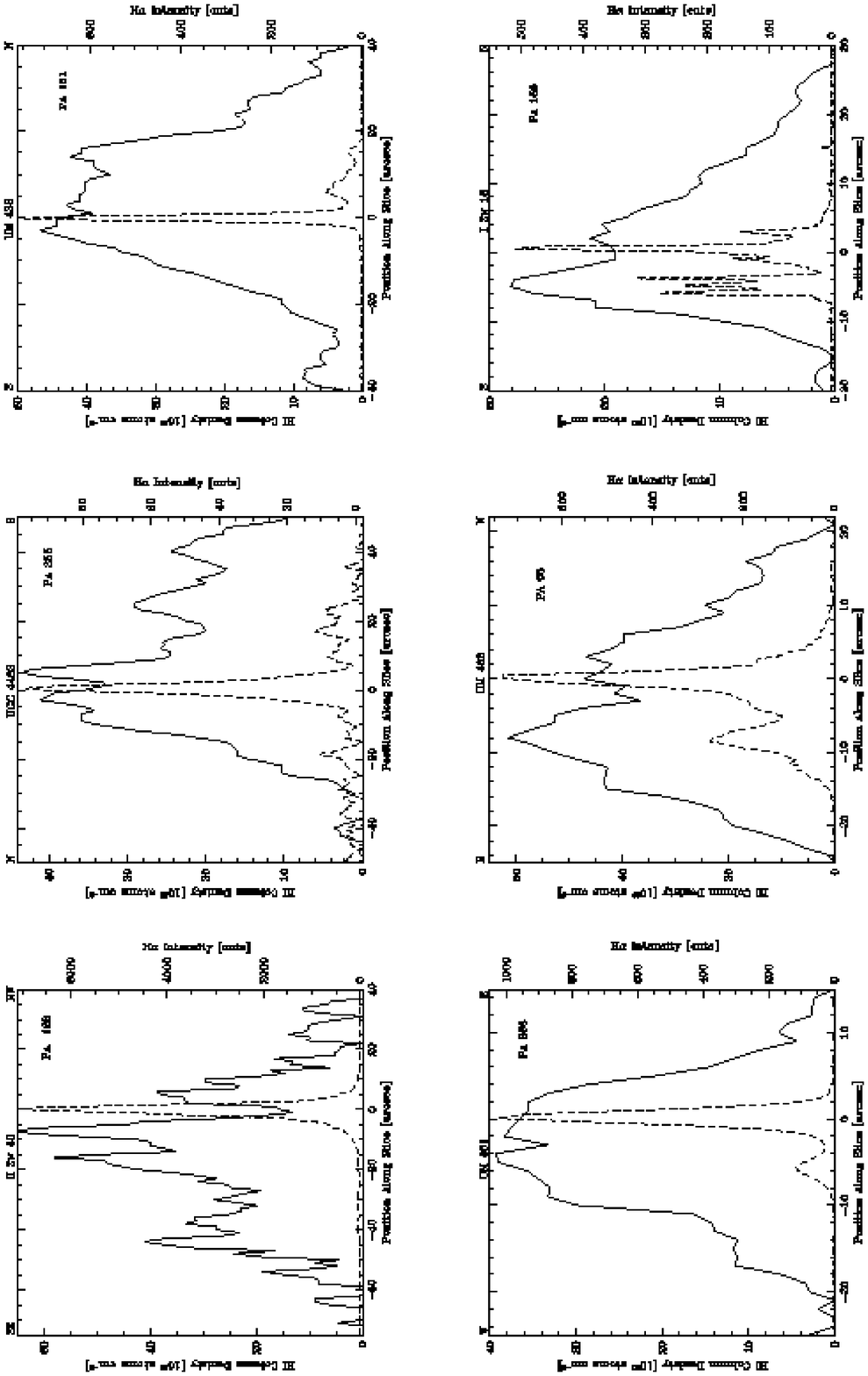,width=6.in,bbllx=30pt,bblly=-50pt,bburx=550pt,bbury=820pt,clip=t}
\vfill
\eject
\figcaption[vanzee.fig14.ps]{ HI and H$\alpha$ intensity profiles.  Each cut is centered on the
peak H$\alpha$ emission feature with the position angle set such that the
slice will intersect additional HII regions.  In each panel, the HI column density 
is denoted by a solid line; the H$\alpha$ flux (in units of counts, not on a common 
scale) is denoted by a dashed line.  The H$\alpha$ resolution is
approximately 2\arcsec~while the HI maps have resolutions of approximately 5\arcsec. 
The data for I~Zw~18 were taken from van Zee \etal (1998).
\label{fig:cuts} }

\vfill
\eject
\psfig{figure=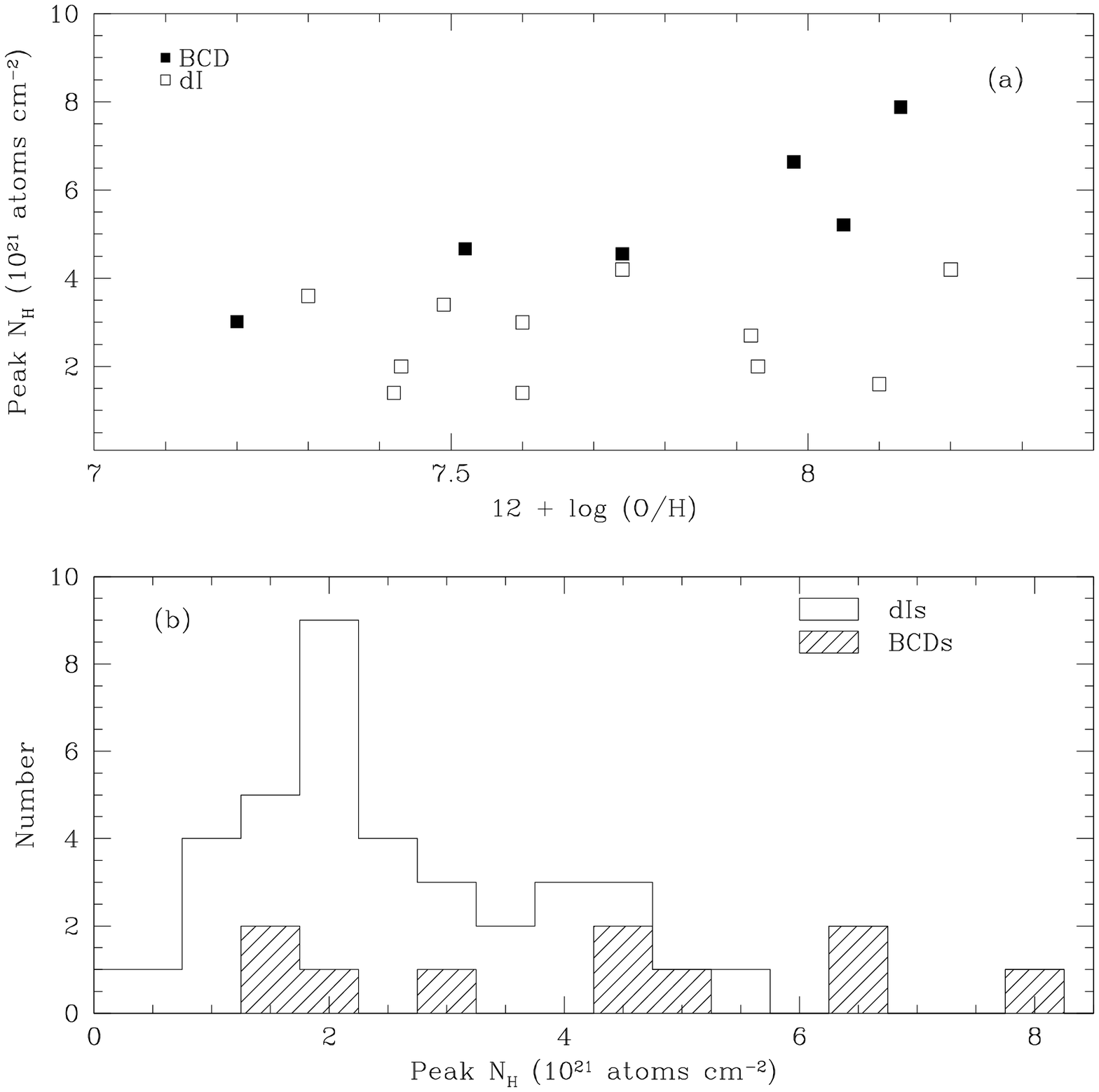,width=6.in,bbllx=1pt,bblly=10pt,bburx=600pt,bbury=700pt,clip=t}
\vskip -1.5 truein
\figcaption[vanzee.fig15.ps]{ (a) Peak column density as a function of metallicity.  The BCD points
(black squares) include I~Zw~18 (van Zee \etal 1998) and the five BCDs in the 
present sample.  The dIs (open squares) are from the samples of Skillman \etal (1988),
Shostak \& Skillman (1989), Lo \etal (1993), and van Zee \etal (1997b).   
The peak column densities have not been corrected for inclination or for beam smearing
effects.   (b) A histogram of peak column densities for BCDs (hatched) and dIs (open). 
The BCD sample includes the present sample, galaxies in Taylor \etal (1994),
and van Zee \etal (1998); the dI sample also includes galaxies in Skillman \etal (1987), C\^ot\'e (1995),  Kobulnicky \& Skillman (1995), and Hunter \etal (1996), as well as 
the references mentioned above.
The column densities
have not been corrected for inclination or beam effects, but it is suggestive that
the BCDs have systematically higher column densities than dIs. \label{fig:column} }

\vfill
\eject
\footnotesize
\tablenum{1}
\begin{deluxetable}{llccccccccccc}
\tabcolsep 2pt
\tablewidth{40pc}
\tablecaption{Global Parameters \label{tab:global}}
\tablehead{
\colhead{} & \colhead{} & \colhead{} & \colhead{} & \colhead{} & \colhead{} & \colhead{} & \colhead{} & \colhead{} & \colhead{} & \colhead{} & \colhead{Log} & \colhead{} \\[.2ex]
\colhead{} &\colhead{Morph.\tablenotemark{a}}&\colhead{}&\colhead{Distance\tablenotemark{b}}& \multicolumn{2}{c}{\underline{ \ \ $D_{25}$\tablenotemark{c} \ \ } }& \colhead{} &\colhead{M$_{\rm HI}$} & \colhead{\underline{D$_{\rm HI}$}\tablenotemark{e}} & \colhead{M$_{\rm HI}$/L$_{\rm B}$}& \colhead{SFR\tablenotemark{f}} & \colhead{(M$_{\rm HI}$/SFR)}  & \colhead{12 +}\\[.2ex]
\colhead{Galaxy}&\colhead{Type}& \colhead{Multi.\tablenotemark{a}} &\colhead{[Mpc]}& \colhead{[\arcsec]}& \colhead{[kpc]} &\colhead{M$_{\rm B}$\tablenotemark{d}} & \colhead{[10$^8$ M$_{\odot}$]} & \colhead{D$_{25}$} & \colhead{[M$_{\odot}$/L$_{\odot}$]}& \colhead{[M$_{\odot}$ yr$^{-1}$]} & \colhead{[yr]} & \colhead{log(O/H)\tablenotemark{g}}}
\startdata
II Zw 40 & BCD I   & double   & 10.1 & 37 & 1.81 & --16.2 & 4.38  & 3.4 & 1.0 & 1.2   & 8.6 & 8.13 $\pm$ 0.15 \nl
UGC 4483 & BCD II  & double   & 3.44 & 67 & 1.12 & --12.7 & 0.372 & 3.4 & 2.0 & 0.0031 & 10.1 & 7.52 $\pm$ 0.03 \nl
UM 439   & BCD II: & multiple & 14.6 & 39 & 2.76 & --15.8 & 3.12  & 4.2 & 1.0 & 0.083 & 9.6 & 8.05 $\pm$ 0.15 \nl
UM 461   & BCD II  & double   & 13.9 & 20 & 1.35 & --14.4 & 1.71  & 4.9 & 1.9 & 0.071 & 9.4 & 7.74 $\pm$ 0.15 \nl
UM 462   & BCD II  & double   & 13.9 & 31 & 2.09 & --16.1 & 2.65  & 2.8 & 0.6 & 0.23  & 9.1 & 7.98 $\pm$ 0.15 \nl
\enddata
\tablenotetext{a}{Classification and multiplicity from Telles {\it et al.} (1997); we have
adopted their classification scheme for UGC 4483.}
\tablenotetext{b}{Distance references: II Zw 40-- Brinks \& Klein (1988); 
UGC 4483-- Tolstoy {\it et al.} (1995); UM 439, UM 461, UM 462-- Taylor {\it et al.} (1995).}
\tablenotetext{c}{Isophotal diameter references: II Zw 40, UM 439, UM 461, UM 462-- Salzer {\it et al.} (1998a);
UGC 4483-- RC3.}
\tablenotetext{d}{Apparent blue magnitude references: II Zw 40, UM 439, UM 461, UM 462-- Telles {\it et al.} (1997); 
UGC 4483-- Stavely--Smith {\it et al.} (1992).}
\tablenotetext{e}{HI Diameter for II Zw 40 measured at a PA of 115.}
\tablenotetext{f}{H$\alpha$ flux references: II Zw 40, UM 439, UM 461, UM 462-- Salzer {\it et al.} (1998a); UGC 4483-- Skillman {\it et al.} (1994)}
\tablenotetext{g}{Oxygen abundance references: II Zw 40, UM 439, UM 461, UM 462-- Melnick {\it et al.} (1988); 
UGC 4483-- Skillman {\it et al.} (1994).}
\end{deluxetable}

\tablenum{2}
\begin{deluxetable}{llcr}
\tablewidth{23pc}
\tablecaption{HI Synthesis Imaging Observing Log \label{tab:vlaobs}}
\tablehead{
\colhead{} &\colhead{}& \colhead{} & \colhead{T$_{\rm int}$} \\[.2ex]
\colhead{Galaxy} &\colhead{Date} & \colhead{Config.} & \colhead{[min]}  }
\startdata
II Zw 40 & 1997 February  4 & BnA & 162.0 \nl
         & 1997 April  4    & B   & 70.5  \nl
         & 1997 June  9     & BnC & 115.5 \nl
         & 1997 June 26     & C   & 57.5  \nl
         & 1997 September 5 & CS  & 39.0  \nl
UGC 4483 & 1997 April  1    & B   & 178.0 \nl
         & 1997 April 11    & B   & 196.0 \nl
         & 1997 June 28     & C   & 175.5 \nl
         & 1997 August 14   & CS  & 71.0  \nl
UM 439   & 1997 April  4    & B   & 378.5 \nl
         & 1997 June 28     & C   & 169.5 \nl
         & 1997 August 14   & CS  & 79.0  \nl
UM 461/2 & 1997 February  4 & BnA & 73.0  \nl
         & 1997 April 11    & B   & 344.5 \nl
         & 1997 June 28     & C   & 173.0 \nl
         & 1997 August 14   & CS  & 76.5  \nl
\enddata
\end{deluxetable}

\tablenum{3}

\begin{deluxetable}{ccccrcc}
\tablewidth{40pc}
\tablecaption{Parameters of the HI Data Cubes \label{tab:maps}}
\tablehead{
\colhead{}&\colhead{} &\colhead{} & \colhead{} & \colhead{}& \colhead{}& \colhead{linear} \\[.2ex]
\colhead{}&\colhead{Robustness}& \colhead{uv taper} & \colhead{uv range} & \colhead{Beam}& \colhead{rms}  & \colhead{resolution} \\[.2ex]
\colhead{Galaxy}  &\colhead{Parameter} &\colhead{[k$\lambda$~k$\lambda$]} & \colhead{[k$\lambda$~k$\lambda$]} & \colhead{[arcsec$\times$arcsec]} & \colhead{[mJy beam$^{-1}$]} & \colhead{[pc beam$^{-1}$]} } 
\startdata
II Zw 40 & 5   & 15~~~15 &  0~~~20 & 17.0~$\times$~15.4 & 0.73 & 830 $\times$ 760 \nl
         & 5   & \nodata & \nodata &  8.6~$\times$~~7.6 & 0.64 & 420 $\times$ 370 \nl
         & 0.5 & \nodata & \nodata &  5.7~$\times$~~4.8 & 0.79 & 280 $\times$ 240 \nl
UGC 4483 & 5   & 15~~~15 &  0~~~20 & 19.3~$\times$~16.1 & 0.71 & 320 $\times$ 270 \nl
         & 5   & \nodata & \nodata & 10.8~$\times$~~8.9 & 0.65 & 180 $\times$ 150 \nl
         & 0.5 & \nodata & \nodata &  7.3~$\times$~~5.3 & 0.72 & 120 $\times$~ 90 \nl
UM 439   & 5   & 15~~~15 &  0~~~20 & 23.3~$\times$~16.3 & 0.78 & 1630 $\times$ 1160 \nl
         & 5   & \nodata & \nodata & 11.0~$\times$~~9.6 & 0.80 & 780 $\times$ 680 \nl
         & 0.5 & \nodata & \nodata &  7.2~$\times$~~5.8 & 0.75 & 510 $\times$ 410 \nl
UM 461/2 & 5   & 15~~~15 &  0~~~20 & 19.6~$\times$~15.5 & 0.59 & 1320 $\times$ 1050 \nl
         & 5   & \nodata & \nodata & 10.1~$\times$~~8.4 & 0.52 & 680 $\times$ 560 \nl
         & 0.5 & \nodata & \nodata &  6.6~$\times$~~5.2 & 0.63 & 440 $\times$ 350 \nl
\enddata
\end{deluxetable}

\tablenum{4}

\begin{deluxetable}{lccccccccc}
\tablewidth{40pc}
\tablecaption{Optical Imaging Observing Log \label{tab:optobs} }
\tablehead{
\colhead{} &\colhead{}&\colhead{}& \colhead{} & \colhead{}& \colhead{FOV} & \colhead{pixel scale} & \colhead{seeing} & \colhead{T$_{\rm int}$} & \colhead{} \\[.2ex]
\colhead{Galaxy} &\colhead{Filter} &\colhead{Date}& \colhead{Observatory}  & \colhead{CCD} &\colhead{[arcmin]} & \colhead{[\arcsec/pix]} & \colhead{[arcsec]} & \colhead{[min]} &\colhead{Observer}}
\startdata
II Zw 40 & R         & Dec 1996 & KPNO 0.9m & T2KA & 11.7 $\times$ 11.7 & 0.69 & 2.6 & 3$\times$5  & LvZ \nl
         & H$\alpha$ & Nov 1989 & KPNO 0.9m & TI 2 &  2.5 $\times$  2.5 & 0.43 & 2.0 & 3$\times$30 & JJS \nl
UGC 4483 & R         & Dec 1996 & KPNO 0.9m & T2KA & 11.7 $\times$ 11.7 & 0.69 & 3.6 & 2$\times$5  & LvZ \nl
         & H$\alpha$ & Mar 1987 & JKT  1.0m & GEC  &  1.9 $\times$  2.9 & 0.30 & 1.8 & 1$\times$15 & EDS \nl
UM 439   & B         & Apr 1990 & KPNO 0.9m & TI 6 &  2.5 $\times$  2.5 & 0.43 & 1.5 & 3$\times$20 & JJS \nl
         & H$\alpha$ & Apr 1990 & KPNO 0.9m & TI 6 &  2.5 $\times$  2.5 & 0.43 & 1.5 & 2$\times$20 & JJS \nl
UM 461   & B         & Apr 1990 & KPNO 0.9m & TI 6 &  2.5 $\times$  2.5 & 0.43 & 1.7 & 3$\times$20 & JJS \nl
         & H$\alpha$ & Apr 1990 & KPNO 0.9m & TI 6 &  2.5 $\times$  2.5 & 0.43 & 1.5 & 2$\times$20 & JJS \nl
UM 462   & B         & Apr 1990 & KPNO 0.9m & TI 6 &  2.5 $\times$  2.5 & 0.43 & 1.9 & 2$\times$20 & JJS \nl
         & H$\alpha$ & Apr 1990 & KPNO 0.9m & TI 6 &  2.5 $\times$  2.5 & 0.43 & 1.5 & 2$\times$20 & JJS \nl
\enddata
\end{deluxetable}

\tablenum{5}

\begin{deluxetable}{lcrcrcccccr}
\tabcolsep 2pt
\tablewidth{40pc}
\tablecaption{Kinematic Parameters \label{tab:hiparms}}
\tablehead{
\colhead{} & \multicolumn{2}{c}{Kinematic Center} & \colhead{Systemic} \\[.2ex]
\colhead{} &  \colhead{RA}&\colhead{Dec}& \colhead{Velocity} &\colhead{$<$PA$>$}&  \colhead{$<$i$>$} & \colhead{R$_{\rm max}$} & \colhead{V$(R_{\rm max})$} &\colhead{M$_{\rm dyn}$}  & \colhead{\underline{M$_{\rm HI}$} }&  \colhead{$<\sigma>$}\\[.2ex]
\colhead{Galaxy} & \colhead{(1950)} & \colhead{(1950)} & \colhead{[km s$^{-1}$]} & \colhead{[deg]} &\colhead{[deg]}&  \colhead{[kpc]} & \colhead{[km s$^{-1}$]} & \colhead{[10$^8$ M$_{\odot}$]} & \colhead{M$_{\rm dyn}$}&\colhead{[km s$^{-1}$]} }
\startdata
II Zw 40\tablenotemark{a} & 05:53:05.1 &   03:22:56 &  792 & 135$\pm$5 & 60$\pm$10 &\nodata &\nodata&\nodata&\nodata&\nodata \nl
UGC 4483 & 08:32:05.2 &   69:57:03 &  158 &  10$\pm$3 & 60$\pm$10 & 1.00 & 22.9 & 1.29 & 0.32 & 7$\pm$2  \nl
UM 439   & 11:34:02.7 &   01:05:39 & 1094 & 152$\pm$3 & 60$\pm$10 & 2.12 & 42.1 & 8.74 & 0.36 & 11$\pm$3 \nl
UM 461   & 11:48:59.4 & --02:05:42 & 1033 & 290$\pm$5 & 30$\pm$10 &\nodata&\nodata&\nodata&\nodata& 11$\pm$2 \nl
UM 462   & 11:50:03.2 & --02:11:36 & 1053 &  55$\pm$3 & 50$\pm$10 & 2.02 & 30.1 & 4.25 & 0.51 & 11$\pm$2 \nl
\enddata
\tablenotetext{a}{The position angle refers to the axis of the tidal features, not the main body.}
\end{deluxetable}

\tablenum{6}

\begin{deluxetable}{lcccccl}
\tabcolsep 1pt 
\tablewidth{30pc}
\tablecaption{Peak Gas Densities \label{tab:colden}}
\tablehead{
\colhead{}& \colhead{N$_{\rm HI}$} & \multicolumn{2}{c}{\underline{ \ \ \ \ \ \ \ \ \ \ \ \ \ $\Sigma_g$\tablenotemark{a} \ \ \ \ \ \ \ \ \ \ \ \ \ }}&  \multicolumn{2}{c}{\underline{ \ \ \ \ \ \ \ \ \ \ \ \ $<\Sigma_c>$  \ \ \ \ \ \ \ \ \ \ \ \ }} &\colhead{resolution}\\[.2ex]
\colhead{Galaxy} & \colhead{[10$^{21}$ cm$^{-2}$]}  & \colhead{[10$^{21}$ cm$^{-2}$]} &
\colhead{[M$_{\odot}$ pc$^{-2}$]}& \colhead{[10$^{21}$ cm$^{-2}$]}&\colhead{[M$_{\odot}$ pc$^{-2}$]}&  \colhead{[pc beam$^{-1}$]} }
\startdata
II Zw 40 & 7.88 & 5.27 & 42.2 & \nodata & \nodata & 280 $\times$ 240 \nl
         & 6.90 & 4.62 & 37.0 &         &         & 420 $\times$ 370 \nl
         & 4.56 & 3.05 & 24.4 &         &         & 830 $\times$ 750 \nl
UGC 4483 & 4.67 & 3.12 & 25.0 &   1.8   &   14.4  & 120 $\times$~ 90 \nl
         & 2.86 & 1.92 & 15.4 &         &         & 320 $\times$ 270 \nl
         & 2.09 & 1.40 & 11.2 &         &         & 640 $\times$ 540 \nl
UM 439   & 5.21 & 3.49 & 27.9 &   2.7   &   21.6  & 510 $\times$ 410 \nl
         & 3.83 & 2.57 & 20.6 &         &         & 780 $\times$ 680 \nl
UM 461   & 4.55 & 5.27 & 42.2 & \nodata & \nodata & 440 $\times$ 350 \nl
         & 3.59 & 4.16 & 33.3 &         &         & 680 $\times$ 565 \nl
UM 462   & 6.64 & 4.45 & 35.6 &   2.0   &   16.0  & 440 $\times$ 350 \nl
         & 5.89 & 3.94 & 31.5 &         &         & 680 $\times$ 565 \nl
\enddata
\tablenotetext{a}{$\Sigma_g$ has been corrected for inclination and neutral helium content.}
\end{deluxetable}

\end{document}